\documentclass[12pt,epsfig]{article}
\pdfoutput=1
\usepackage{jheppub}
\usepackage{graphicx}  

\usepackage{dcolumn}   
\usepackage{bm}        
\usepackage{amssymb}   

\newcommand{\beq}{\begin{equation}}
\newcommand{\eeq}[1]{\label{#1}\end{equation}}
\def\beqa{\begin{eqnarray}}
\def\eeqa#1{\label{#1}\end{eqnarray}}
\newcommand{\eeqn}{\end{equation}}
\newcommand{\CR}{\notag \\}
\newcommand{\leqn}[1]{(\ref{#1})}

\def\sw{s_W}
\def\cw{c_W}
\def\mhu{m_{H_u}}
\def\mhd{m_{H_d}}
\def\tb{t_\beta}

\def\stacksymbols #1#2#3#4{\def\theguybelow{#2}
    \def\vp{\lower#3pt}
    \def\sp{\baselineskip0pt\lineskip#4pt}
    \mathrel{\mathpalette\intermediary#1}}

\def\intermediary#1#2{\vp\vbox{\sp
     \everycr={}\tabskip0pt
     \halign{$\mathsurround0pt#1\hfil##\hfil$\crcr#2\crcr
              \theguybelow\crcr}}}

\def\gsim{\stacksymbols{>}{\sim}{2.5}{.2}}
\def\lsim{\stacksymbols{<}{\sim}{2.5}{.2}}
\begin{document}

\title{XENON100 Implications for Naturalness in the MSSM, NMSSM and $\lambda$-SUSY}

\author{Maxim Perelstein} 
\author{and Bibhushan Shakya}

\affiliation{Laboratory of Elementary Particle Physics, 
	     Cornell University, Ithaca, NY 14853, USA}
	     
\emailAdd{mp325@cornell.edu} \emailAdd{bs475@cornell.edu}     

\abstract{In a recent paper arXiv:1107.5048, we discussed the correlation between the elastic neutralino-nucleon scattering cross section, constrained by dark matter direct detection experiments, and fine-tuning at tree-level in the electroweak symmetry breaking sector of the Minimal Supersymmetric Standard Model (MSSM). Here, we show that the correlation persists in the Next-to-Minimal Supersymmetric Standard Model (NMSSM), and its variant, $\lambda$-SUSY. Both models are strongly motivated by the recent discovery of a 125 GeV Higgs-like particle. We also discuss the implications of the recently published bound on the direct detection cross section from 225 live days of XENON100 experiment. In both the MSSM and the NMSSM, most of the parameter space with fine-tuning less than 10\% is inconsistent with the XENON100 bound. In $\lambda$-SUSY, on the other hand, large regions of completely natural electroweak symmetry breaking are still allowed, primarily due to a parametric suppression of fine-tuning with large $\lambda$. The upcoming XENON1T experiment will be able to probe most of the parameter space with less than 1\% fine-tuning in all three models.
}


\maketitle

\newpage

\section{Introduction}

Several experiments around the world are currently attempting to observe dark matter via ``direct detection", {\it i.e.} measuring recoils of detector nuclei following their collisions with ambient dark matter particles. While no convincing observation has been reported so far, the sensitivity of the experiments is rapidly increasing. Currently, the best upper bounds on the cross section of elastic, spin-independent dark matter-nucleon scattering in the 10 GeV-TeV mass range come from the XENON100 experiment~\cite{xenon100paper,xenon2012}, and are of the order $10^{-44}-10^{-45}$ cm$^2$. This is the range where the predictions of many attractive theoretical models of dark matter lie~\cite{DMreviews}. The most studied of these is R-parity conserving supersymmetry, where the lightest neutralino $\chi^0_1$ generically has the right properties to explain dark matter. It is therefore important to understand the implications of the direct detection bounds for supersymmetric dark matter. 

In a recent paper~\cite{ftmssm}, we pointed out a strong correlation between direct detection cross sections and naturalness of electroweak symmetry breaking (EWSB) in the Minimal Supersymmetric Standard Model (MSSM)\footnote{Related discussions have also appeared in Refs.~\cite{others,zeptobarn}.}: MSSM parameter points with lower direct detection cross section have more finely-tuned EWSB. This conclusion seems very general: It does not depend on the details of the SUSY-breaking mechanism and high-scale physics, and for most of the parameter space (except pure-Higgsino dark matter) it does not require imposing the thermal relic density constraint, so that it applies even in models with non-standard early cosmological history. The only assumption is the absence of accidental cancellations among physically distinct contributions to the dark matter-nucleon scattering amplitudes ({\it e.g.} Higgs- and squark-exchange diagrams). 

The main goal of the present paper is to extend this study to the Next-to-Minimal Supersymmetric Standard Model (NMSSM). (For reviews, see, for example, Refs.~\cite{mainref,NMSSMreview}.). Our main motivation to study the NMSSM is the recent discovery of a new particle, with a mass of about 125 GeV and properties consistent with the Standard Model (SM) Higgs, at the LHC~\cite{LHC_Higgs}. In the MSSM, a Higgs particle of this mass can only be accommodated at a price of severe fine-tuning in the EWSB~\cite{HPR}. Fine-tuning is significantly alleviated if an extra singlet (with respect to SM gauge groups) superfield is coupled to the Higgs sector, as in the NMSSM~\cite{HPR}. In this respect, a particularly promising variation of the NMSSM is the ``$\lambda$-SUSY"~\cite{lambdasusy},  characterized by a large superpotential coupling $\lambda$ between the singlet and doublet Higgs fields. The addition of a singlet has a non-trivial effect on dark matter phenomenology due to the possible admixture of the singlino in $\chi^0_1$, as well as the additional SM-singlet Higgs state. Does the correlation between direct-detection rates and EWSB fine-tuning persist in the NMSSM and $\lambda$-SUSY? It should be noted that in \cite{ftmssm} and in this paper, we are only interested in the tree-level naturalness of the $Z$ mass. Of course, additional fine-tuning may be induced by the radiative corrections associated with heavy stops and gluinos; in this sense, the fine-tuning measure we use should be interpreted as the lower bound on the total fine-tuning. We are not aware of any correlation between the fine-tuning due to radiative corrections and dark matter direct detection cross sections.

This paper is organized as follows. We begin by reviewing and updating the MSSM results of Ref.~\cite{ftmssm} in Section~\ref{sec:review}, including the latest results from XENON100~\cite{xenon2012}. We then briefly review the structure of the NMSSM and $\lambda$-SUSY in Section~\ref{sec:NMSSM}. Section~\ref{sec:FT} provides definitions of fine-tuning, and discusses a suppression of fine-tuning at large values of $\lambda$, which will be important for understanding our results. The analysis procedure is outlined in Section~\ref{sec:scan}, and the results are presented and discussed in Section~\ref{sec:results}. Finally, we conclude in Section~\ref{sec:conc} with a brief summary of the main results and directions for future work.

\section{Naturalness and Direct Detection in the MSSM: a Mini-Review and Update}
\label{sec:review}

\begin{figure}[t]
\begin{center}
\centerline {
\includegraphics[width=4.5in]{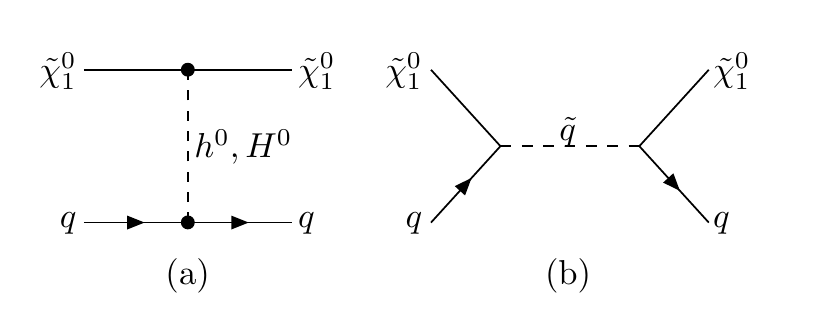}
}
\caption{Feynman diagrams contributing to spin-independent elastic scattering of neutralino dark matter off a nucleon in the MSSM.}
\label{fig:Fd1}
\end{center}
\end{figure}

Spin-independent scattering of the neutralino off a quark in the MSSM is mediated by $t$-channel Higgs or $s$-channel squark exchanges, Fig.~\ref{fig:Fd1}. As explained in Ref.~\cite{ftmssm}, any cancellation between the two contributions should be regarded as accidental. We assume that no such accidental cancellations occur, so that the total cross section is {\it at least} of the same order as the contribution of either diagram class by itself. Since we are interested in the lowest direct detection cross section possible for a fixed amount of fine-tuning, it is sufficient for our purposes to examine a single diagram class. We focus on the Higgs exchange diagrams, since they contain the parameters that directly affect electroweak symmetry breaking, and we ignore the squark exchanges\footnote{Recent LHC limits also suggest that the squarks are heavy, making the contribution from the squark exchange diagrams subdominant.}; this gives a cross-section that is less than or equal to the total cross-section, which is sufficient for the purpose of this paper. The cross section, ignoring loop corrections, then depends on just five MSSM parameters: $\mu$, $M_1$, $M_2$, $\tan\beta$, and $m_A$. The same parameters suffice to determine the degree of fine-tuning in the EWSB at tree level, as well as the composition of the lightest supersymmetric particle (LSP). This allows us to study correlations among these quantities without any assumptions about the nature of the SUSY breaking sector, high-scale physics such as unification, etc. Such a study was performed in Ref.~\cite{ftmssm}, and the reader interested in the details of the analysis is referred to that paper. Here, we summarize the conclusions and update the main plots of Ref.~\cite{ftmssm} to include the newly released XENON100 cross section bound from 225 days of running in 2010-2012~\cite{xenon2012}. We also use the updated values of nuclear form factors, which enter the direct detection cross section predictions (from~\cite{nucs}): 
\beq
f_{Tu}^{(p)} = 0.023,~~f_{Td}^{(p)} = 0.033,~~f_{Ts}^{(p)} = 0.26.
\eeq{fnum}
It should be kept in mind that there is significant uncertainty in the values of the form factors, especially that of the strange quark, which plays a dominant role for a typical MSSM point.\footnote{We refer to~\cite{zeptobarn} for a more extended discussion of the effects of varying the strange quark form factor. A recent discussion of the determination of this form factor can be found in Ref.~\cite{formfactordiscussion}. } In particular, lattice QCD results~\cite{lattice} suggest a lower value $f_{Ts}^{(p)} \sim 0.05$. Correlations among physical quantities ({\it e.g.} direct detection cross section and fine-tuning) are independent of the particular value of $f_{Ts}^{(p)}$. However, it does play a role in the interpretation of cross section bounds, introducing an order-one uncertainty. We will return to this issue in Section~\ref{sec:results}.

\begin{figure}[t]
\centering
\includegraphics[width=3.0in,height=2.1in]{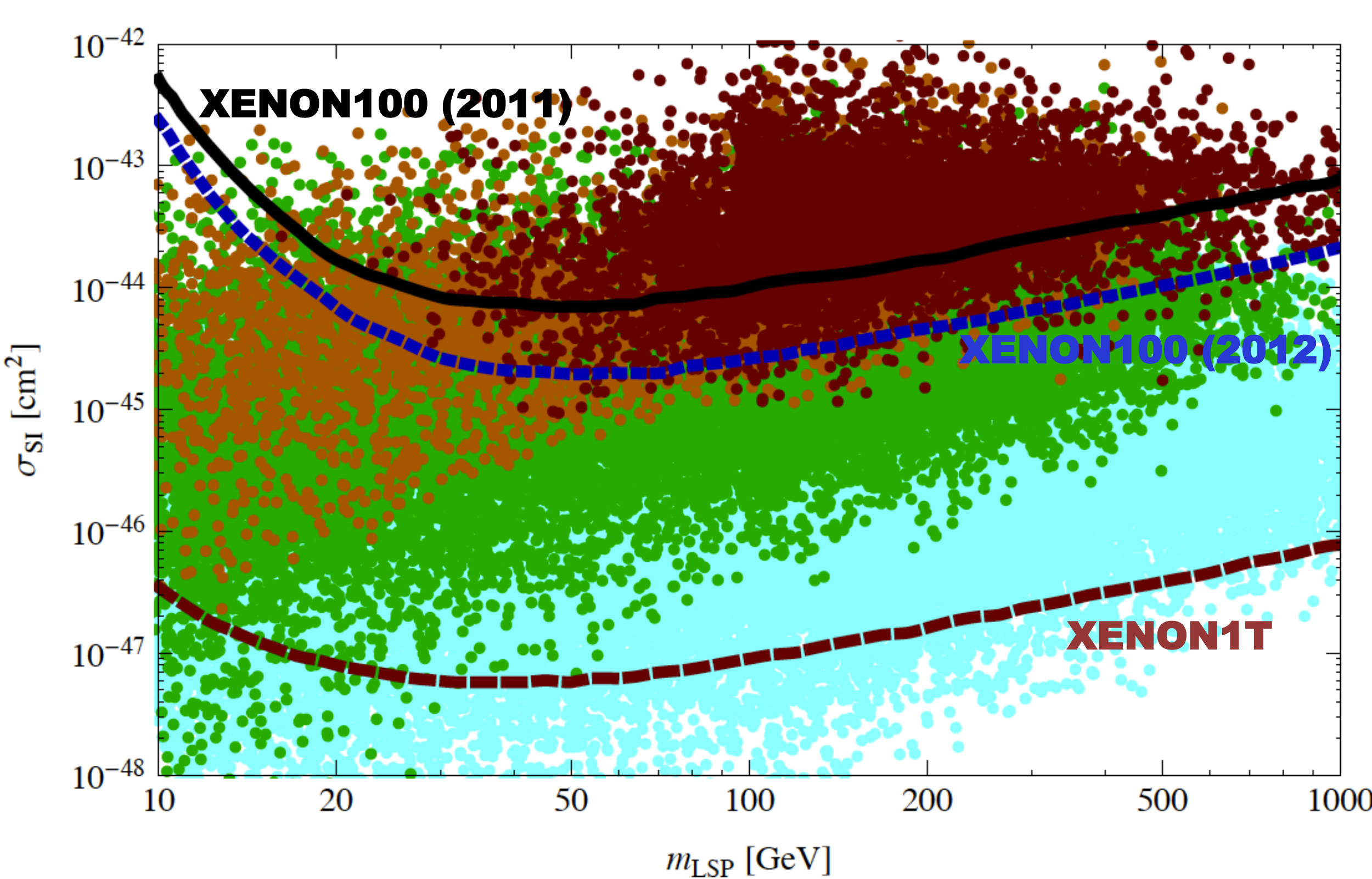}
\includegraphics[width=3.0in,height=2.1in]{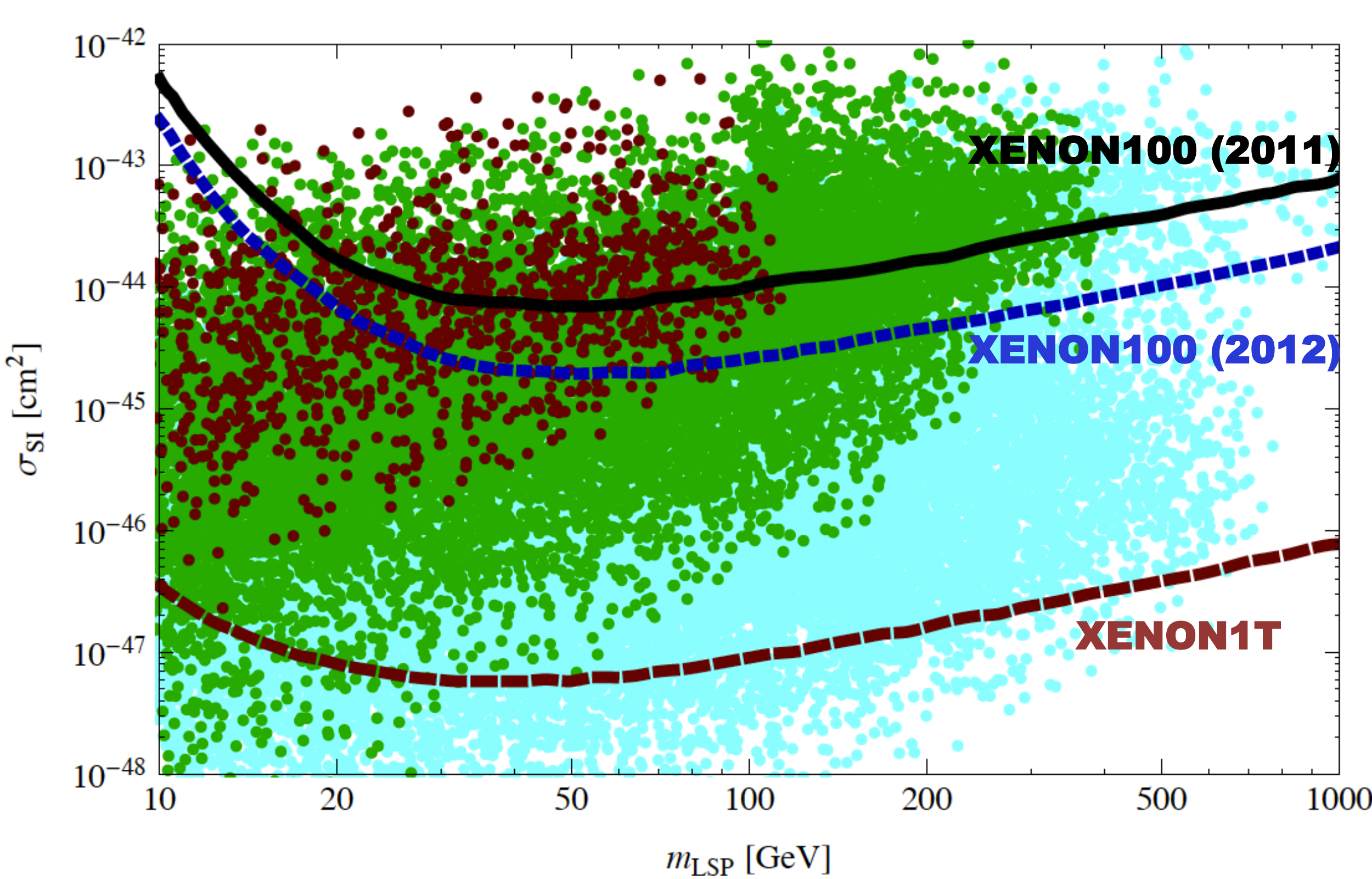}
\caption{Left panel: Direct detection cross section vs. the LSP mass, for MSSM points with purity above 0.2 (red), between 0.1 and 0.2 (orange), 0.01 and 0.1 (green), and 10$^{-3}$ and 0.01 (cyan). Right panel: Direct detection cross section vs. the LSP mass, for MSSM points with gaugino-like LSP. Red, green and cyan points correspond to EWSB fine-tuning in the intervals $(0, 10)$; $[10, 100)$; $[100,1000)$, respectively. The black/solid and blue/dotted lines correspond to the XENON100 100 days/2011~\cite{xenon100paper} and 225 days/2012~\cite{xenon2012} exclusion limits, respectively. The red/dashed line shows the projected sensitivity of XENON1T~\cite{xenon1T}.  Real values of the scanned MSSM parameters are assumed, and points with strong accidental cancellations have been discarded.}
\label{fig:MSSM}
\end{figure}

The main results of Ref.~\cite{ftmssm} are summarized by the two plots in Fig.~\ref{fig:MSSM}. Making use of the LHC Higgs data now available, these results now include m$_h=126$ GeV and constraints on CP-odd Higgs decays to $\tau$ pairs from CMS \cite{CMS:gya}. The MSSM parameters scanned over are the same as in Ref.~\cite{ftmssm} (with mass parameters allowed to have either sign):
\beqa
& &M_1 \in [10, 10^4]~{\rm GeV};~~~~~M_2 \in [80, 10^4]~{\rm GeV}; \CR
& &\mu \in [80, 10^4]~{\rm GeV};~~~~~m_A \in [100, 10^4]~{\rm GeV}; \CR
& & \tan\beta \in [2,50]\,.
\eeqa{scanbound_1}

 First, there is a correlation between the direct detection cross section and the ``purity" of the LSP, defined as $p=\min(F_H, 1-F_H)$, where $F_H$ is the Higgsino fraction. A typical mixed gaugino/Higgsino LSP has a direct detection cross section around $10^{-44}$ cm$^2$; significantly lower cross sections are only possible for pure gaugino or pure Higgsino states. There is a direct correlation between the cross section and purity of the LSP: Lower cross sections require higher purity. The left panel of Fig.~\ref{fig:MSSM} shows that almost no points with purity above 0.2 are allowed by the 2012 XENON100 bound. Second, in the case of predominantly gaugino LSP, there is a correlation between purity and fine-tuning of EWSB: suppressing the Higgsino admixture requires raising $\mu$, which in turn increases fine-tuning. (A more detailed discussion of fine-tuning will be given in Section~\ref{sec:FT}.) Thus, in the gaugino LSP region, points with direct detection cross sections significantly below $10^{-44}$ cm$^2$ are necessarily fine-tuned, and the smaller the cross section, the more severe the required fine-tuning is. The right panel of Fig.~\ref{fig:MSSM} shows this correlation (points with greater than 90\% higgsino fraction have been removed), and the implications of XENON100. Very few points with fine-tuning better than 1/10 survive the 2012 XENON100 constraint, and the majority of those have LSP masses below 50 GeV. XENON-1T will be able to probe fine-tuning down to 1\% level for most of the relevant mass range. 
 
 The effect of including constraints on the CP-odd Higgs decays to $\tau$ pairs is clearly seen in these figures; a majority of the points with cross section above $10^{-43}$ cm$^2$ are now eliminated (compare to the corresponding figures in \cite{ftmssm}). This correlation arises because points with enhanced decays to $\tau$s inconsistent with the CMS data are associated with large $\tan\beta$ (corresponding to large down-type Yukawas), which also give large direct detection cross-sections since the Higgs couplings to the strange  and down quarks, which provide the major nuclear contents, are likewise enhanced. 

All above statements apply without assuming that the relic density of neutralinos is set by thermal decoupling, and so are remarkably insensitive to cosmological evolution assumptions. In the Higgsino region, there is no correlation between direct detection cross section and fine-tuning. However, if one further assumes standard cosmology and thermal decoupling, the correct relic density for Higgsinos requires $\mu \sim 1$ TeV, corresponding to EWSB fine-tuning of about 1/500, independent of the direct detection bounds. The bounds depend strongly, however, on the local dark matter density, which is assumed to be $0.3\,$GeVcm$^{-3}$; recent studies indicate that this value might be larger (see e.g. \cite{localdensity}), which would strengthen the bounds discussed here. 

\section{The NMSSM and $\lambda$-SUSY}
\label{sec:NMSSM}

The NMSSM is obtained by adding to the MSSM a gauge-singlet superfield $\hat{S}$, with a superpotential containing
\begin{equation}
\lambda \hat{S}\hat{H_u}\cdot\hat{H_d}+\frac{\kappa}{3}\hat{S}^3\,.
\end{equation}
This is the most general superpotential consistent with a ${\cal Z}_3$ symmetry under which $\hat{S}$, $\hat{H}_u$ and $\hat{H}_d$ fields have charge $1/3$. The $\mu \hat{H}_u\hat{H}_d$ term present in the MSSM is not allowed by this symmetry. 
The soft SUSY-breaking Lagrangian contains the following new terms:
\begin{equation}
m_S^2|S|^2+(\lambda A_\lambda H_u\cdot H_d S+\frac{1}{3}\kappa A_\kappa S^3+~{\rm h.c.})\,,
\end{equation}
while the $B\mu H_u H_d$ term of the MSSM is forbidden by the ${\cal Z}_3$ symmetry.
When the singlet field receives a vacuum expectation value (vev), $s=\langle S\rangle$, after SUSY breaking, effective $\mu$ and $B$ terms are generated:
\begin{equation}
\mu_{eff}=\lambda s,~~~~~B_{eff}=A_\lambda+\kappa s\,.
\end{equation}
The two bosonic components of $S$ give one CP-even and one CP-odd Higgs bosons, which mix with the MSSM Higgses to produce 3 CP-even and 2 CP-odd mass eigenstates. The extended Higgs sector is described in terms of the following parameters:
\begin{equation}
\label{parameters}
\lambda,\kappa,A_\lambda,A_\kappa, m^2_{H_u},m^2_{H_d},m^2_{S}\,.
\end{equation}
Of these, $\lambda$ is particularly important in accommodating a 125 GeV Higgs, since it enters the tree-level mass of the CP-even Higgses. Large values of $\lambda$ are needed to raise the Higgs mass. Renormalization group evolution drives $\lambda$ to larger values at high energies, and eventually $\lambda$ hits a Landau pole. Requiring this pole to lie above the conventional grand unification (GUT) scale leads to an upper bound on the weak-scale value of $\lambda$, 
\beq
\lambda\lsim 0.75\,.
\eeq{NMSSM_lambda}
In this paper, we refer to the theory with $\lambda$ obeying this condition as the ``NMSSM". However, it is also interesting to consider the regime of larger $\lambda$. Even though the simple prediction of gauge coupling unification is lost, such theories provide perfectly consistent descriptions of physics at the weak scale, and require the least fine-tuning among models of this class to incorporate the 125 GeV Higgs~\cite{HPR}. Avoiding a Landau pole below 10 TeV (which would almost certainly be inconsistent with precision electroweak constraints) requires
\beq
0.75 \lsim \lambda \lsim 2.0\,.
\eeq{lSUSY_lambda}
In this study we will refer to the theory with $\lambda$ in this range as the ``$\lambda$-SUSY"~\cite{lambdasusy}. Above the 10 TeV scale, $\lambda$-SUSY needs to be incorporated into a more fundamental ultraviolet (UV) theory; see, for example, Refs.~\cite{naturalsusy,MR} for recent attempts at UV model building in this context.

The fermionic component of $S$, the ``singlino", mixes with the four neutralinos of the MSSM to give 5 neutralino mass eigenstates $\chi^0_i$. The lightest of these, $\chi^0_1$, is assumed to be the LSP and is stable due to conserved R-parity. We will assume that all of dark matter present in the Earth's neighborhood consists of the $\chi^0_1$ particles. The neutralino mass matrix has the form
\beq
M_{\chi^0}= \left( \begin{tabular}{ccccc} $M_1$ & $0$ & $-m_Z \sw \cos\beta$ & $m_Z \sw \sin\beta$ & $0$\\ 
                                                                        $0$ & $M_2$ & $m_Z \cw \cos\beta$ & $-m_Z \cw \sin\beta$ & $0$\\ 
                                                     $-m_Z \sw \cos\beta$ & $m_Z \cw \cos\beta$ & $0$ & $-\mu_{eff}$ & $-\lambda v \sin\beta$\\ 
                                                     $m_Z \sw \sin\beta$ & $-m_Z \cw \sin\beta$ & $-\mu_{eff}$ & $0$ & $-\lambda v \cos\beta$\\ 
                                       $0$ & $0$ & $-\lambda v \sin\beta$ & $-\lambda v \cos\beta$ & $2\frac{\kappa}{\lambda} \mu_{eff}$\\ \end{tabular} \right)\,,
\eeq{neu_masses}
where $\sw$ and $\cw$ are the sine and cosine of the Weinberg angle, and $\tan\beta=v_u/v_d$. The only two new parameters that enter, beyond those listed in~\leqn{parameters}, are the weak-ino soft masses $M_1$ and $M_2$.  

\subsection{Direct Detection}

The spin-independent LSP-nucleon scattering in the NMSSM and $\lambda$-SUSY occurs via the same diagrams as in the MSSM, see Fig.~\ref{fig:Fd1}, except now all three CP-even Higgses can be exchanged in the $t$-channel. We again ignore the squark diagram, with the same motivation as in the MSSM study. The scattering amplitude can be easily computed using the NMSSM Feynman rules listed, for example, in Ref.~\cite{NMSSMreview}; in particular, the LSP-Higgs coupling is given by
\begin{eqnarray}
\label{coupling}
g(h_i\tilde{\chi}^0_1\tilde{\chi}^0_1) &=& \frac{\lambda}{\sqrt{2}}(s_{H_d}n_{\tilde{H}_u} n_{\tilde{S}} + s_{H_u}n_{\tilde{H}_d}n_{\tilde{S}}+s_Sn_{\tilde{H}_u}n_{\tilde{H}_d})-\frac{\kappa}{\sqrt{2}}\,s_Sn_{\tilde{S}}n_{\tilde{S}}\nonumber\\
&&+\frac{g_1}{2}(s_{H_d}n_{\tilde{B}}n_{\tilde{H}_d}-s_{H_u}n_{\tilde{B}}n_{\tilde{H}_u})
-\frac{g_2}{2}(s_{H_d}n_{\tilde{W_0}}n_{\tilde{H}_d}-s_{H_u}n_{\tilde{W_0}}n_{\tilde{H}_u})\,,
\end{eqnarray} 
where $s_\alpha$ and $n_\beta$ denote the relevant components of the $i$-th Higgs mass eigenstate ($i=1\ldots 3$) and the lightest neutralino, respectively. The cross section is then obtained using the standard formalism~\cite{DMreviews}; we use the values of nuclear form factors listed in~\leqn{fnum}. 

\section{Fine-Tuning in MSSM, NMSSM and $\lambda$-SUSY}
\label{sec:FT}

The supersymmetric parameters must produce the weak scale; this is the origin of potential fine-tuning, reflected in the tree-level relation for the $Z$ boson mass:
\beq
\frac{1}{2}m_Z^2=\frac{m^2_{H_d}-\tan^2\beta~m^2_{H_u}}{\tan^2\beta-1}-\mu^2\,.
\eeq{mz}
This relation holds in the MSSM as well as in the NMSSM and $\lambda$-SUSY. A reasonable expectation is that $\mu$ and $m_{H_u}$ must also be around the $m_Z$ scale, and some fine-tuning is required to obtain the correct $m_Z$ if these terms are far above this scale. A quantitative measure of the amount of fine-tuning is the variation in $m_Z^2$ resulting from variations of the fundamental Lagrangian parameters of the theory. This can be defined as 
\begin{equation}
\Delta=\text{max}~\Delta_i\,,~~~\Delta_i \equiv \left|\frac{\partial~\text{log}~m_Z^2}{\partial~\text{log}~\xi_i}\right|\,,
\end{equation}
where the index $i$ runs over all independent parameters in the Lagrangian. 
There is an ambiguity over the scale at which the parameters $\xi_i$ are defined; a common choice in the literature is the GUT scale~\cite{BG}. We will instead use $\xi_i$ defined at the {\it weak} scale, since we are working with general (N)MSSM without assuming specific SUSY-breaking scenarios. At tree-level, the only relevant parameters in the MSSM are $\mu$, $\mhu$, $\mhd$ and $b$; analytic expressions for $\Delta_i$, $i=1\ldots 4$ are given in Refs.~\cite{PS,ftmssm}. 

In the NMSSM, minimization of the Higgs potential relates the scalar vevs $v_u$, $v_d$ and $s$ to the Lagrangian parameters~\cite{mainref, ftmeasure}:
\begin{eqnarray}
\nonumber E_1\equiv m_{H_u}^2+\mu^2+\lambda^2v_d^2+\frac{g^2}{2}(v_u^2-v_d^2)-\frac{v_d}{v_u}\mu(A_\lambda+\kappa s)=0\,, \nonumber\\
 E_2\equiv m_{H_d}^2+\mu^2+\lambda^2v_u^2+\frac{g^2}{2}(v_d^2-v_u^2)-\frac{v_u}{v_d}\mu(A_\lambda+\kappa s)=0 \,,\nonumber\\
E_3\equiv  m_{S}^2+\frac{\kappa}{\lambda}A_\kappa \mu+2\frac{\kappa^2}{\lambda^2}\mu^2+\lambda^2 (v_u^2+v_d^2) -2\lambda\kappa v_u v_d-\lambda^2v_uv_d\frac{A_\lambda}{\mu}=0\,,
\label{E1E3}
\end{eqnarray}
where 
\begin{equation}
m_Z^2=g^2v^2,~~~~~v\equiv \sqrt{v_u^2+v_d^2} = 174~\text{GeV},~~~~~\mu=\lambda s\,.
\end{equation}
We defined $g^2=(g_1^2+g_2^2)/2 \approx 0.52$, where $g_1$ and $g_2$ are the SM $U(1)_Y$ and $SU(2)_L$ couplings respectively. 

Recall that all parameters are defined independently at the weak scale\footnote{It is possible that a cancellation that appears finely tuned from the weak-scale point of view may be rendered natural in a particular high-scale theory; this would be missed by our approach. See, for example, Ref.~\cite{Antusch:2012gv}.}. Because of this, the effects of RG running from a high scale down to the weak scale are not captured; in particular, large corrections to $m^2_{H_u}$ in $E_1$ from heavy stops, the traditional source of fine-tuning for a heavy Higgs, are not included, and the setup implemented in this paper explores a different source of fine-tuning. 

The variations of $m_Z$ under input parameter changes can be calculated in the following way~\cite{ftmeasure}: Imposing that the minimization conditions~\leqn{E1E3} continue to hold under variations of the input parameters, one obtains
\begin{equation}
\label{pv}
\delta E_j=\sum_i\frac{\partial E_j}{\partial \xi_i}\delta \xi_i+\frac{\partial E_j}{\partial m_Z^2}\delta m_Z^2+\frac{\partial E_j}{\partial \tan\beta}\delta\tan\beta+\frac{\partial E_j}{\partial \mu}\delta\mu=0\,,
\end{equation}
for $j=1\ldots 3$, where $i$ runs over the fundamental parameters listed in Eq.~\leqn{parameters}. These three equations can be solved for $\delta m_Z^2,~\delta\tan\beta,$ and $\delta\mu$. Defining
\begin{equation}
\frac{\partial E_j}{\partial \xi_i}=P_{ij},~~~\frac{\partial E_j}{\partial m_Z^2}=Z_j,~~~ \frac{\partial E_j}{\partial \tan\beta}=T_j,~~~ \frac{\partial E_j}{\partial \mu}=M_j\,,
\end{equation}
we obtain
\begin{equation}
\Delta_i\equiv \left| \frac{\partial\,\text{log}\,m_Z^2}{\partial\,\text{log}\,\xi_i} \right|= \left|\frac{\xi_i}{m_Z^2}\frac{\delta m_Z^2}{\delta \xi_i}\right|=\left|-\frac{\xi_i}{m_Z^2}\frac{\sum_{jkl}\epsilon^{jkl}~P_{ij}T_kM_l}{\sum_{jkl}\epsilon^{jkl}~Z_jT_kM_l}\right|\,.
\label{delta-NMSSM}
\end{equation}
The fine-tuning is then calculated by taking the maximum of these $\Delta_i$'s, as described earlier. 

\subsection{Fine-Tuning Suppression in $\lambda$-SUSY}
\label{sec:FTsup}

Since we define fine-tuning as sensitivity with respect to weak-scale input parameters, no renormalization group evolution is involved in computing it, and the formalism described above for the NMSSM applies equally well to $\lambda$-SUSY. However, there is a very interesting, and potentially important, parametric suppression of fine-tuning at $\mu\gg m_Z$ and $\mhu\gg m_Z$ in $\lambda$-SUSY regime that is not present in the NMSSM. We will discuss this phenomenon in this subsection.

In the MSSM, the sensitivity of $m_Z$ to $\mu$ is given simply by
\beq
\Delta_\mu = \frac{4\mu^2}{m_Z^2}\,,
\eeq{Delta_mu}
so that fine-tuning scales as $(\mu/m_Z)^2$ at large $\mu$, independent of all other parameters. Likewise, large values of $\mhu$ give fine-tuning of order $(\mhu/m_Z)^2$.
In the NMSSM and $\lambda$-SUSY, the situation is more complicated, since $\mu$ is not an input parameter, but an output of the minimization of the scalar potential. In particular, cancellations between terms in Eq.~\leqn{mz} may occur {\it naturally}, if the minimization conditions force a particular relation of $\mu$ to $\mhu$ and/or $\mhd$. This is exactly what happens, quite generically, in the regime $\lambda\gg g$: the fine-tuning at large $\mu$ and $\mhu$ is parametrically suppressed by a factor of $(g/\lambda)^2$. This is extremely relevant for the LHC searches for supersymmetry: Large $\lambda$ allows one to raise $\mhu$ without increased fine-tuning; the stop masses, which control the size of the quantum correction to $\mhu$, can then be raised by roughly the same amount. If $\lambda/g\sim 3$, as is possible in $\lambda$-SUSY, stop masses in excess of 1 TeV can be completely natural~\cite{HPR}. 

To understand the parametric suppression of fine-tuning at large $\lambda$, consider the model in the particularly simple limit, $A_\lambda=A_\kappa=0$. In this limit, the minimization conditions~\leqn{E1E3} can be solved analytically (see Appendix A for details). The result is
\beqa
\mu^2 &=& \,\mhu^2 \,f_1 \left( \frac{\mhd^2}{\mhu^2}, \frac{m_S^2}{\mhu^2}, \frac{\kappa}{\lambda}\right)\,,\CR
m_Z^2 &=& \,\frac{g^2}{\lambda^2} \, \mhu^2 \,f_2 \left(  \frac{\mhd^2}{\mhu^2}, \frac{m_S^2}{\mhu^2}, \frac{\kappa}{\lambda}\right)\,,
\eeqa{A0mins} 
where corrections of higher orders in $g^2/\lambda^2$ have been ignored, and the functions $f_1$ and $f_2$ are given in Appendix A. These functions are parametrically ${\cal O}(1)$ if all their arguments are order-one numbers, so that generically,
\beq
m_Z^2 \sim \frac{g^2}{\lambda^2} \mhu^2 \sim \frac{g^2}{\lambda^2} \mu^2\,,
\eeq{mzmu}
making the large-$\lambda$ suppression of fine-tuning manifest.

For general $A_\lambda$ and $A_\kappa$, no analytic solution can be found, but there is still a simple way to argue that the suppression of fine-tuning at large $\lambda$ should persist. Consider $\lambda$ dependence of the parameter sensitivities $\Delta_i$ in Eq.~\leqn{delta-NMSSM}. While $T_i$ and $M_i$ appear in both numerator and denominator, $P_{ij}$ appears only in the numerator and $Z_i$ only in the denominator, so the magnitude of $\Delta_i$ is set roughly by the size of $\xi_iP_{ij}$'s relative to the $m_Z^2Z_j$'s. All $Z_j$'s contain a factor ${\lambda^2}/{g^2}$, a term not found in the $P_{ij}$'s, hence $\Delta_i$'s are suppressed by $(g/\lambda)^2$. 

\begin{figure}[t]
\centering
\includegraphics[width=3.0in,height=2.1in]{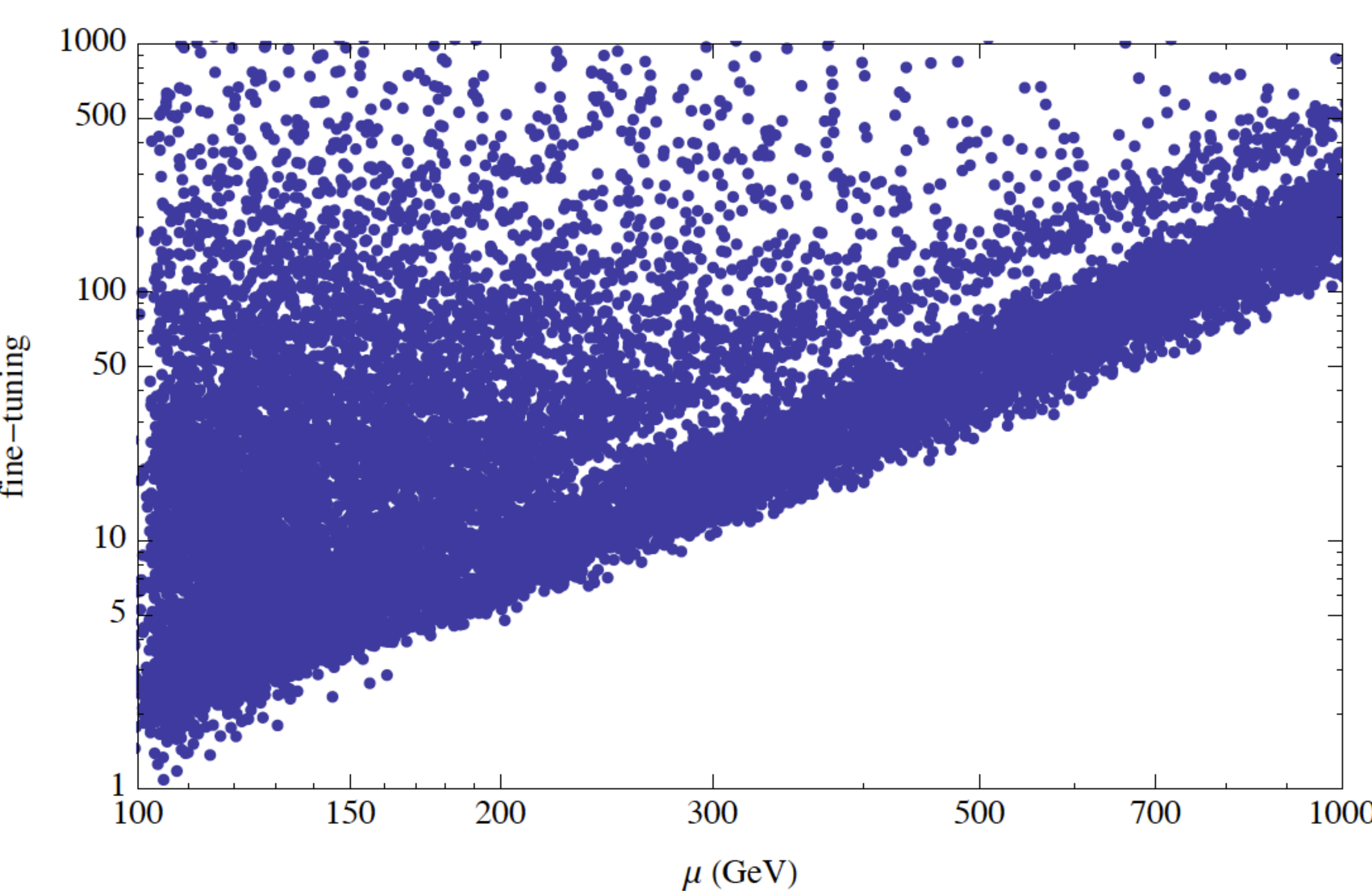}
\includegraphics[width=3.0in,height=2.1in]{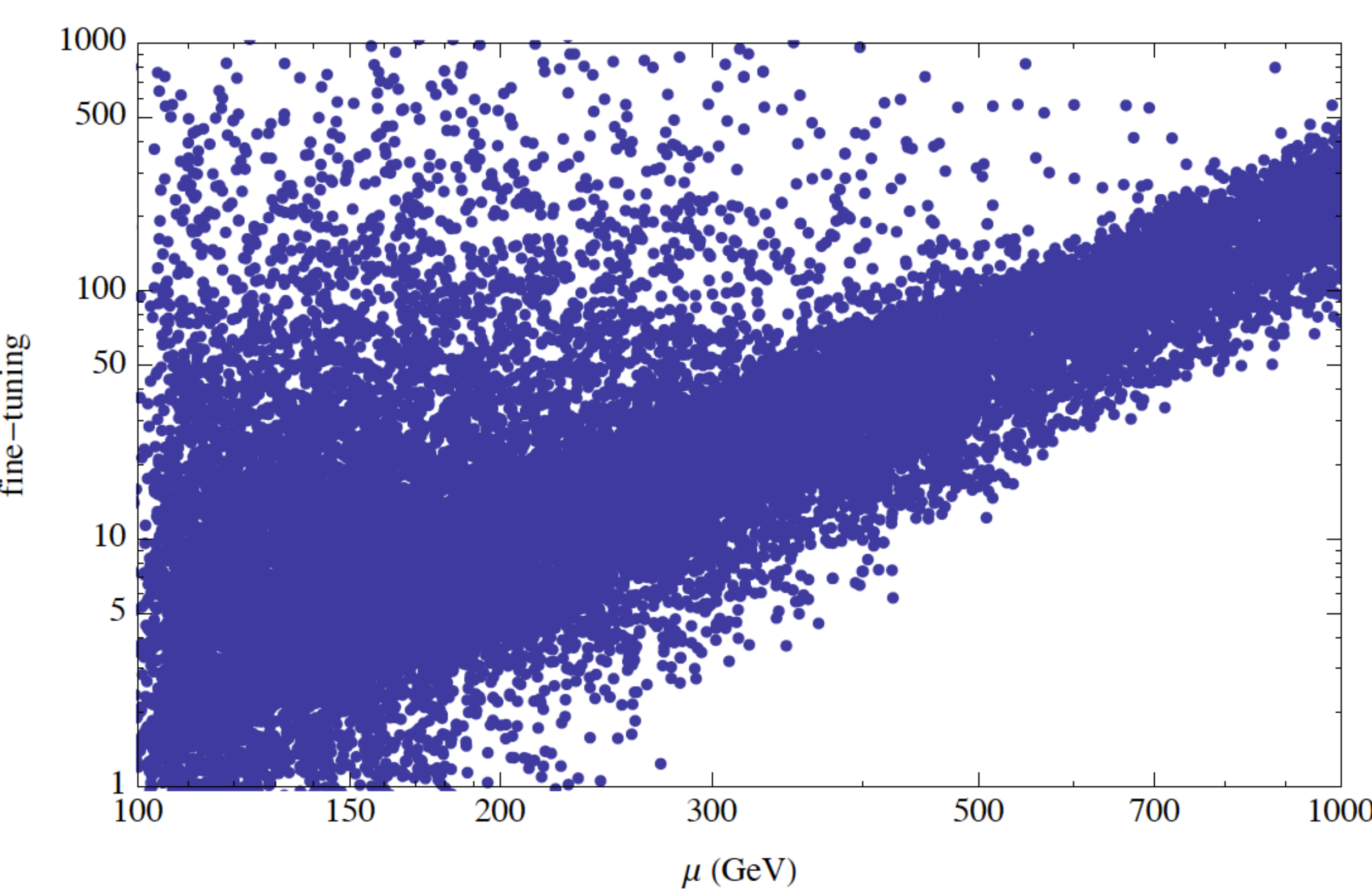}
\caption{Fine-tuning vs $|\mu|$ in the NMSSM (left) and $\lambda$-SUSY (right).}
\label{fig:ftmu}
\end{figure} 

To further illustrate this phenomenon, we plot $\mu$ vs. the amount of fine-tuning in the NMSSM and $\lambda$-SUSY in Figure~\ref{fig:ftmu}. The coupling $\lambda$ at each point varies randomly between 0.4 and 0.75 in the NMSSM scan, and between 0.75 and 2.0 in $\lambda$-SUSY scan. Other details of the scans (which will also be used to study direct detection/fine-tuning correlation) are described in Section~\ref{sec:scan}. Side-by-side comparison of the plots clearly shows that lower fine-tuning for the same value of $\mu$ is possible in $\lambda$-SUSY, consistent with the arguments given above.

\section{Analysis Setup}
\label{sec:scan}

To study correlations between direct detection cross section, fine-tuning and other physical quantities, we performed a scan over the NMSSM/$\lambda$-SUSY parameters. The fine-tuning and direct detection cross section (ignoring the squark contribution and all loop corrections) are completely determined by the 7 Higgs-sector parameters listed in Eq.~\leqn{parameters} and the weak-ino soft masses $M_1$ and $M_2$. For convenience, we used Eqs.~\leqn{E1E3} to interchange $m^2_{H_u},m^2_{H_d},$ and $m_S^2$ with tan$\beta$, $\mu$, and $m_Z^2$. With $m_Z$ fixed, there are 8 parameters to be scanned over. The boundaries of the scan are as follows (all dimensionful parameters are in GeV and can have either sign):

\begin{itemize}
\item $\mu$:~~(70, 10000).\\
Lower values of $\mu$ result in light charginos, failing the LEP-2 bound~\cite{PDG}. Points with $\mu$ above 10 TeV are fine-tuned at the $10^{-3}$ level or worse.  

\item $\lambda$:~~(0.4, 0.75) in the NMSSM scan, (0.75, 2.0) in $\lambda$-SUSY scan.\\
The lower bound is chosen to get a sizable F-term contribution to the tree level Higgs mass, necessary to obtain a 125 GeV Higgs without significant fine-tuning.

\item tan$\,\beta$:~~~(1.7, 54).\\
The upper limit is set by the perturbativity bound on the bottom Yukawa coupling~\cite{Cerdeno:2007sn}. Although $\tan \beta > 54$ is allowed in $\lambda$-SUSY, at large $\tan \beta$, the benefits of having a large value for $\lambda$ are lost. The main motivations for considering $\lambda$-SUSY are (i) a large tree level contribution to the Higgs mass, which helps make a $125$ GeV Higgs more natural, and (ii) a factor of $(\lambda/g)^2$ improvement in EWSB fine-tuning from top/stop loops, compared with the MSSM with the same stop mass. At large $\tan \beta > 54$, both benefits are lost: the F-term contribution to the Higgs mass scales as $1/\tan^2\beta$, while fine-tuning in EWSB goes up with $\tan \beta$. Thus, this model offers no improvement in naturalness, compared to the MSSM, for large $\tan\beta$, which motivated not scanning over those regions. Hence this region, although possible, is not considered, and instead the same upper limit as in the NMSSM is used.

\item $\kappa$:~~~$\pm(10^{-5}, 0.65).$\\
The upper bound is the perturbativity bound in the NMSSM ({\it i.e.}, $\kappa$ remains perturbative up to the GUT scale). 
Since the singlino mass term is $m_S=2\kappa\mu/\lambda$, the region $\kappa<10^{-5}$ typically results in very light ($\textless\,10\,$GeV) singlino LSPs, which are not constrained by XENON100 and will not be considered here. As with  tan$\,\beta$, the upper limit can be extended in $\lambda$-SUSY but does not offer any clear improvements in EWSB fine-tuning, hence will not be considered.

\item $A_\lambda,A_\kappa$:~~(1, 10000). 

\item $M_1$:~~(10, 10000),~~~$M_2$:~~(80, 10000).

\end{itemize}

We generate points randomly distributed with uniform weight in $\lambda, \kappa$, and tan$\beta$, and randomly distributed with logarithmic weight in the other five (dimensionful) parameters. We impose the following constraints:

\begin{itemize}

\item The lightest chargino must be heavier than 103 GeV to satisfy the LEP-2 bound~\cite{PDG}, and must be heavier than the lightest neutralino.

\item The LSP mass must be above 10 GeV (since the current XENON100 data is not sensitive to lower LSP masses).

\item No tachyonic Higgses in the CP even or CP odd sector. (All Higgs masses are computed at tree level.)

\item There is at least one CP-even Higgs with a tree-level mass between 100 and 150 GeV. Fits to the Higgs branching ratios indicate that the observed Higgs is very SM-like: the doublet sector is very close to the decoupling limit, while the singlet fraction of the observed Higgs cannot be too significant. We impose these constraints in our scans in a simple way by requiring that the heavier (non SM-like) doublet and singlet fractions of the Higgs with mass close to 125 GeV be less than 0.1 and 0.5 respectively. This approximately captures the constraints; since our analysis is done at tree level, a more sophisticated fit to individual branching ratios would not be very meaningful.

\item The LSP contribution to the invisible width of the $Z$ must be less than one standard deviation of the measured neutrino contribution: $\Gamma_{Z\rightarrow\chi\chi}\,\textless \,4.2\,$MeV if $m_\chi \,\textless\, m_Z/2$~\cite{PDG,Hooper:2002nq}. The light LSP candidates that survive this constraint are mostly bino. 

\item The CP-odd Higgses must be consistent with CMS bounds on decays to $\tau$ pairs \cite{CMS:gya}.

\end{itemize}

After imposing these conditions, about 25,000 points each remain in the NMSSM and $\lambda$-SUSY scans. All plots in the following section are based on these data sets.

Note that we {\it do not} demand correct thermal relic density of the neutralino for the points in our scan. There are two reasons for this. First, relic density depends on many model parameters ({\it e.g.} squark and slepton masses) in addition to the ones we scan over, and in most cases we expect that it can be ``fixed" by an appropriate choice of those parameters, with no effect on EWSB fine-tuning\footnote{ For instance, in our scans, the natural points with over-produced thermal relic densities have a significant bino component, and the relic density is very sensitive to slepton masses. Making sleptons light reduces the relic density by enabling coannihilations with the sleptons as well as enhancing the sleptons-mediated annihilation into leptons. For these points, the correct relic density can generally be obtained for some appropriate slepton masses. Calculating the relic density for these points would require scanning over slepton masses as well as time-consuming numerical computations, which are only tangentially relevant to the main idea of the paper, hence we avoid doing this.}. Second, not demanding that the LSP be a thermal relic gives our results broader applicability, including scenarios with non-thermal dark matter production, non-standard cosmological evolution, {\it etc.} Only in one special case (pure-Higgsino LSP) shall we consider the effect of including the relic density constraint. 

In discussing the correlations between the direct detection cross section and other quantities, it is important to distinguish parameter points where the direct detection amplitude is strongly suppressed for accidental reasons. A quantitative measure of an accidental cancellation is the dependence of the cross section on the scan parameters: An abnormally strong dependence indicates an accidental cancellation. We will use the ``accidentality", defined as
\beq
{\rm Acc}  \equiv {\rm max}~\left|\frac{\partial~\text{log}~\sigma_{SI}}{\partial~\text{log}~\xi_i}\right|,
\eeq{acc_def} 
to quantify the presence of accidental cancellations at our scan points. This is analogous to our definition of fine-tuning.

\section{Results and Discussion}
\label{sec:results}

\begin{figure}[t]
\centering
\includegraphics[width=3.0in,height=2.1in]{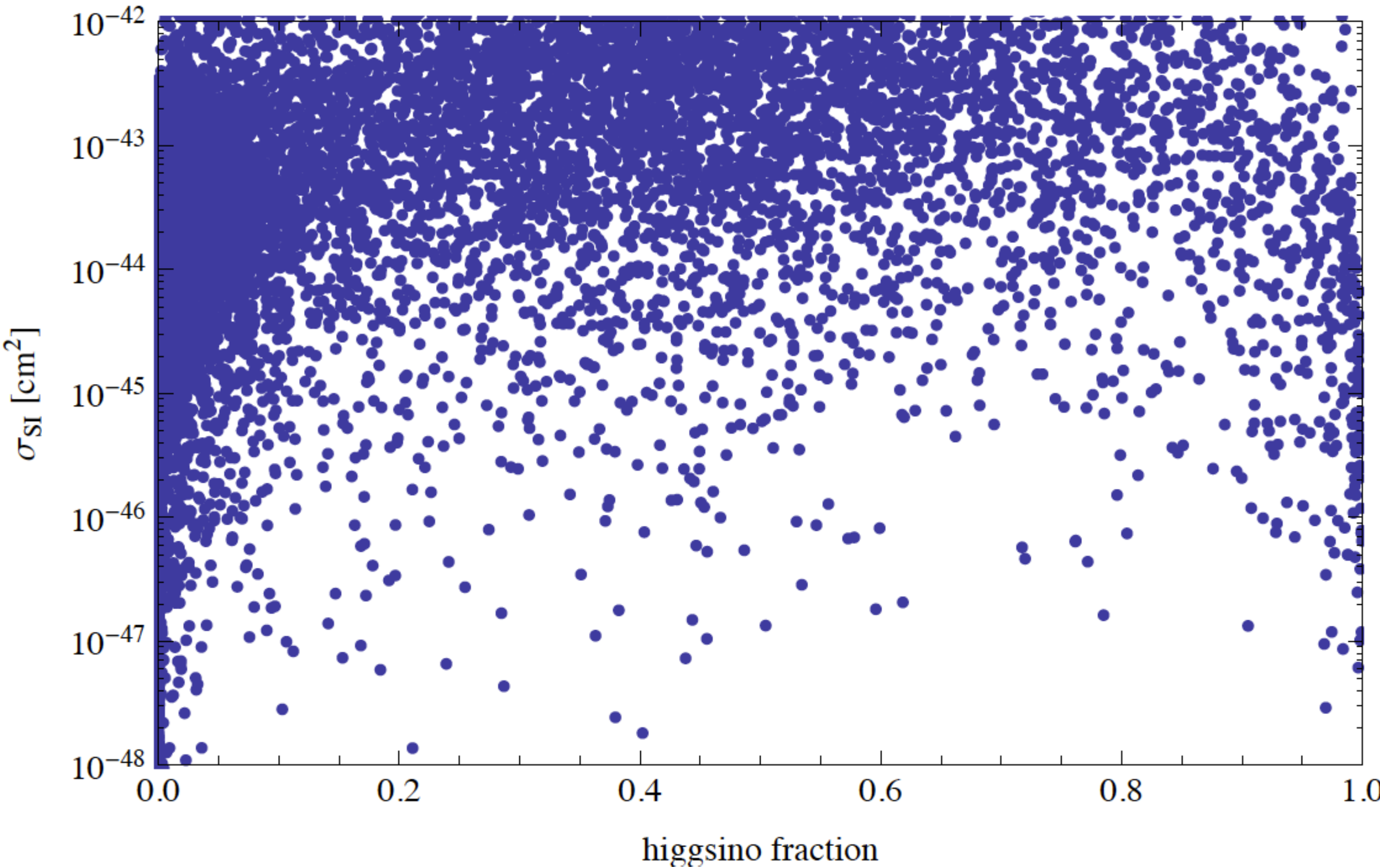}
\includegraphics[width=3.0in,height=2.1in]{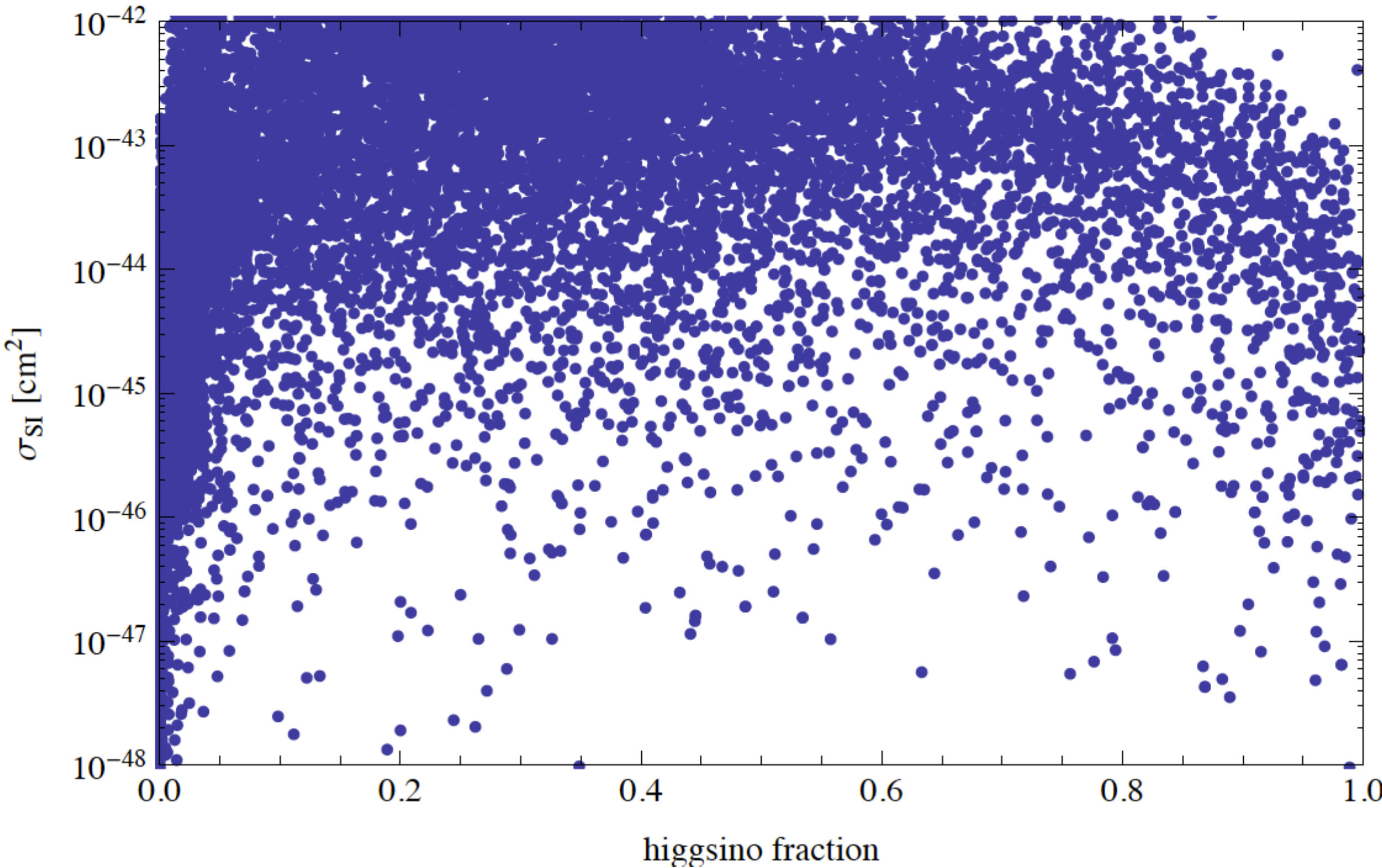}
\\
\includegraphics[width=3.0in,height=2.1in]{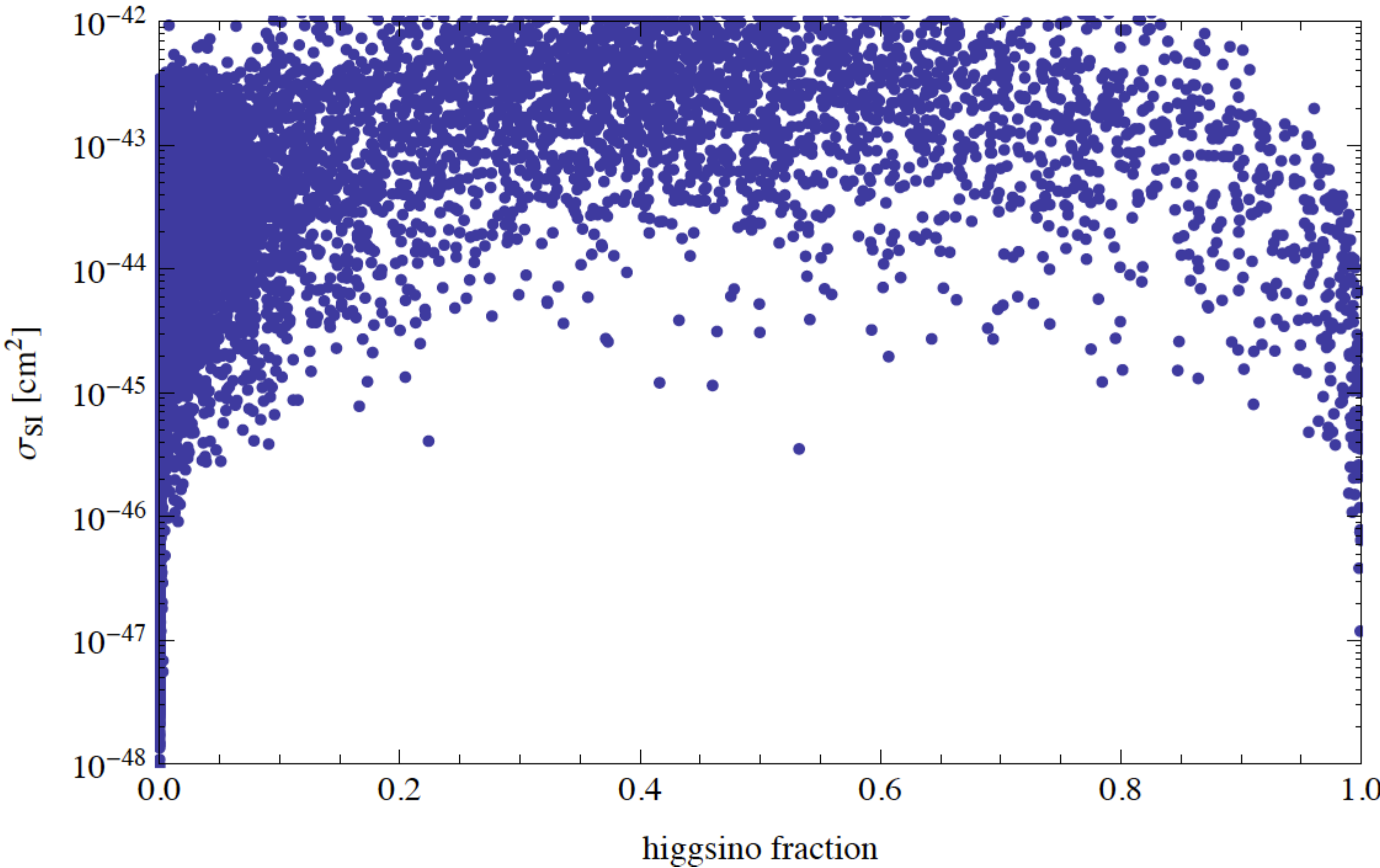}
\includegraphics[width=3.0in,height=2.1in]{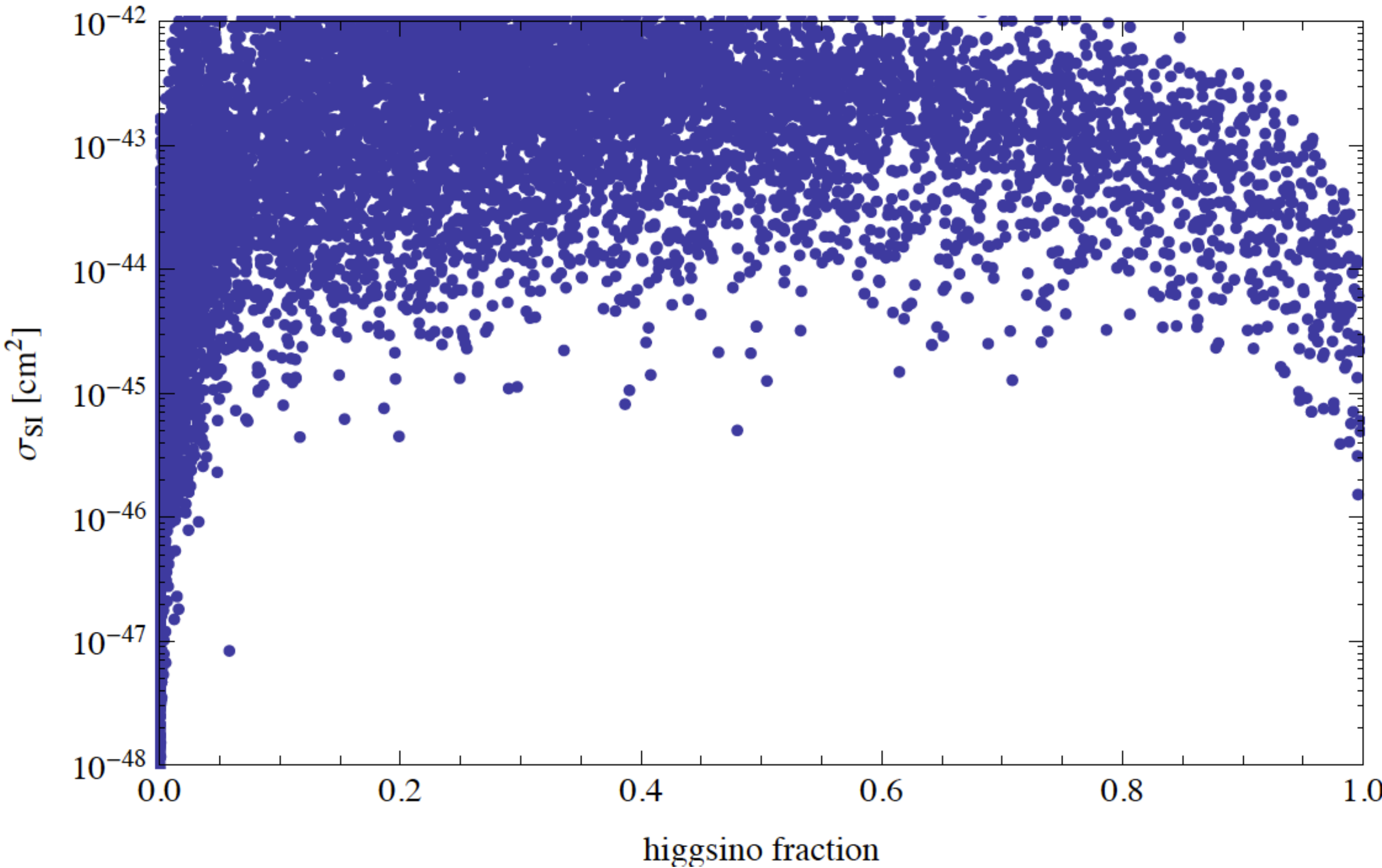}
\caption{Direct detection cross section vs. Higgsino fraction of the LSP, in the NMSSM (left) and $\lambda$-SUSY (right). In the top row, all points are included; in the bottom row, points with accidental cancellations (Acc$>50$) are discarded.}
\label{fig:Hfraction}
\end{figure} 

\begin{figure}[t]
\centering
\includegraphics[width=3.0in,height=2.1in]{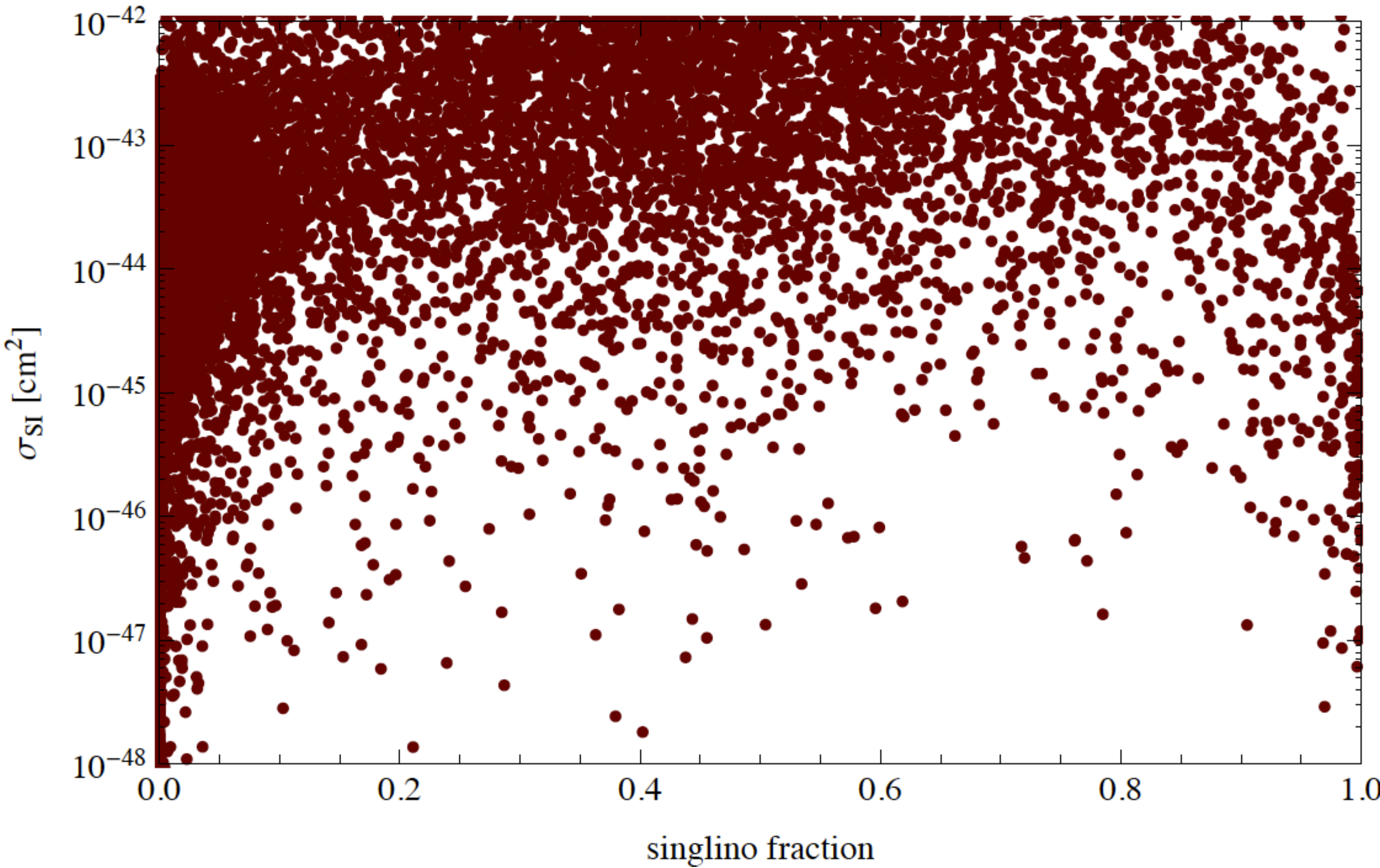}
\includegraphics[width=3.0in,height=2.1in]{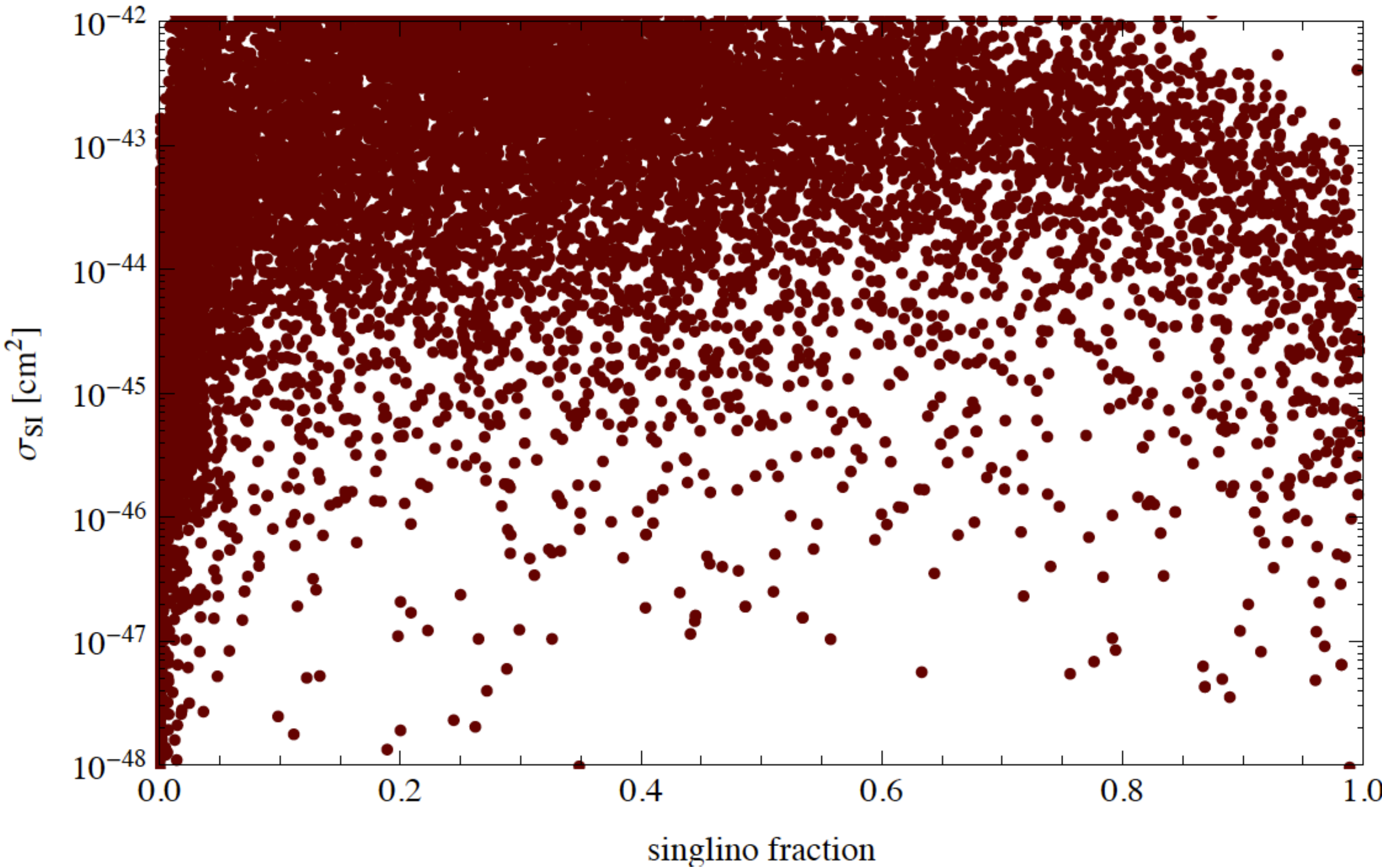}
\\
\includegraphics[width=3.0in,height=2.1in]{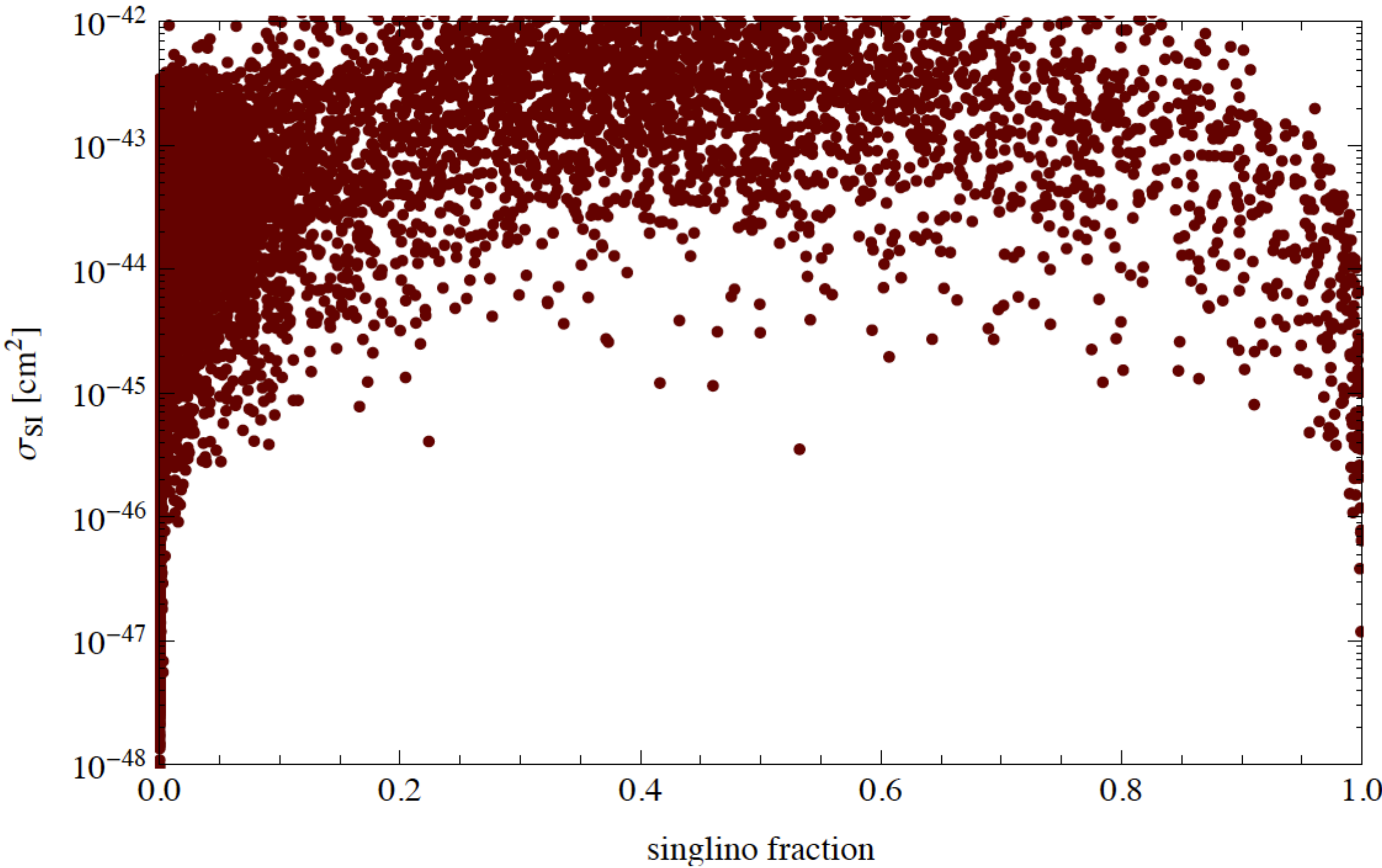}
\includegraphics[width=3.0in,height=2.1in]{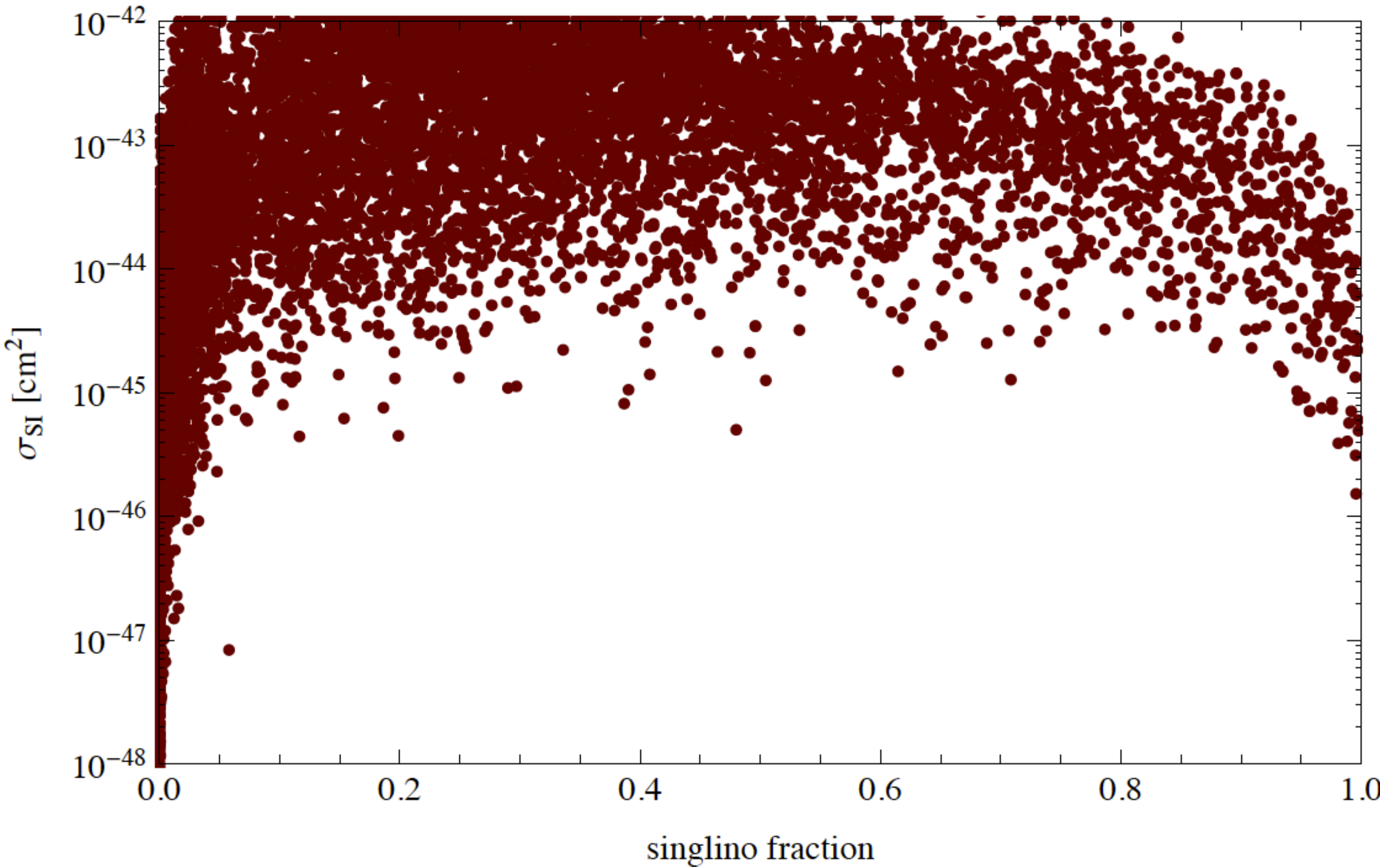}
\caption{Direct detection cross section vs. singlino fraction of the LSP, in the NMSSM (left) and $\lambda$-SUSY (right). In the top row, all points are included; in the bottom row, points with accidental cancellations (Acc$>50$) are discarded.}
\label{fig:Sfraction}
\end{figure} 

First, consider the correlation between the direct detection cross section and the composition of the LSP. This is illustrated in Figs.~\ref{fig:Hfraction} and~\ref{fig:Sfraction}. As in the MSSM, an LSP with a ``generic" composition ({\it i.e.} roughly equal mix of Higgsino, singlino and gaugino) is predicted to have a cross section of about $10^{-45}$ cm$^2$ or higher. This is precisely the region now probed by XENON100. Points with cross sections below $10^{-45}$ cm$^2$ either have accidental cancellations in the cross section, or have an LSP with the Higgsino fraction close to either 1 (``pure Higgsino" regime) or 0 (``pure gaugino/singlino" regime). This can be easily understood by examining the LSP-LSP-Higgs coupling in Eq.~\leqn{coupling}. Schematically, the four terms in the coupling have the form $\tilde{H}\tilde{S}$, $\tilde{H}\tilde{W}$, $\tilde{H} \tilde{B}$, and $\tilde{S}\tilde{S}$. The contributions of the first three terms are clearly suppressed in either pure Higgsino or pure gaugino/singlino limits. The fourth term is not suppressed in the pure-singlino limit, but its contribution is proportional to the coupling $\kappa$ and to the singlet-doublet Higgs mixing angle, both of which can be small. This qualitative behavior is not sensitive to $\lambda$, and thus applies equally in the NMSSM and $\lambda$-SUSY regimes. 

The dependence of the cross section on the singlino fraction of the LSP is qualitatively similar: cross sections below the ``generic" $10^{-45}$ cm$^2$ level occur for singlino fractions close to 0 or 1, see Fig.~\ref{fig:Sfraction}. In principle, small cross sections can also occur for an LSP with order-one singlino and gaugino fractions, and small Higgsino admixture. 
However, such points require a near-degeneracy of the singlino and gaugino (bino or wino) diagonal terms in the neutralino mass matrix, Eq.~\leqn{neu_masses}. This is because there is no direct singlino-gaugino mixing entry in the mass matrix, so the mixing must occur via Higgsinos, which however are nearly decoupled in this region, and can only lead to large mixing if the corresponding diagonal entries are nearly degenerate. Such points rarely occur in the scan, and typically fail the accidentality cut, which explains their absence in the lower panel of Fig.~\ref{fig:Sfraction}.

It is worth pointing out that the direct detection cross section can be suppressed arbitrarily if the LSP is a pure singlino\footnote{We only consider $m_{LSP}\,\textgreater\,10\,$GeV; see e.g.\cite{mainref},\cite{Gunion:2010dy},\cite{Djouadi:2008uj},\cite{Barger:2005hb},\cite{Belikov:2010yi},\cite{Kappl:2010qx} and references therein for discussions of singlino LSP below this mass and related phenomenology.}, and $\lambda$ and $\kappa$ are both small. In this region, however, adding a singlet to the MSSM loses its strongest motivation, since small $\lambda$ suppresses the tree-level contribution to the Higgs mass, and incorporating the 125 GeV Higgs becomes as difficult as in the MSSM. For this reason, we restricted our scan to $\lambda\geq 0.4$, and the small-$\lambda$ region does not appear in our plots. In addition, in $\lambda\to 0$, $\kappa\to 0$ limit the singlino LSP is completely decoupled, so it is difficult to imagine a production mechanism, thermal or non-thermal, that would give it the observed relic abundance. 

\begin{figure}[t]
\centering
\includegraphics[width=3.0in,height=2.1in]{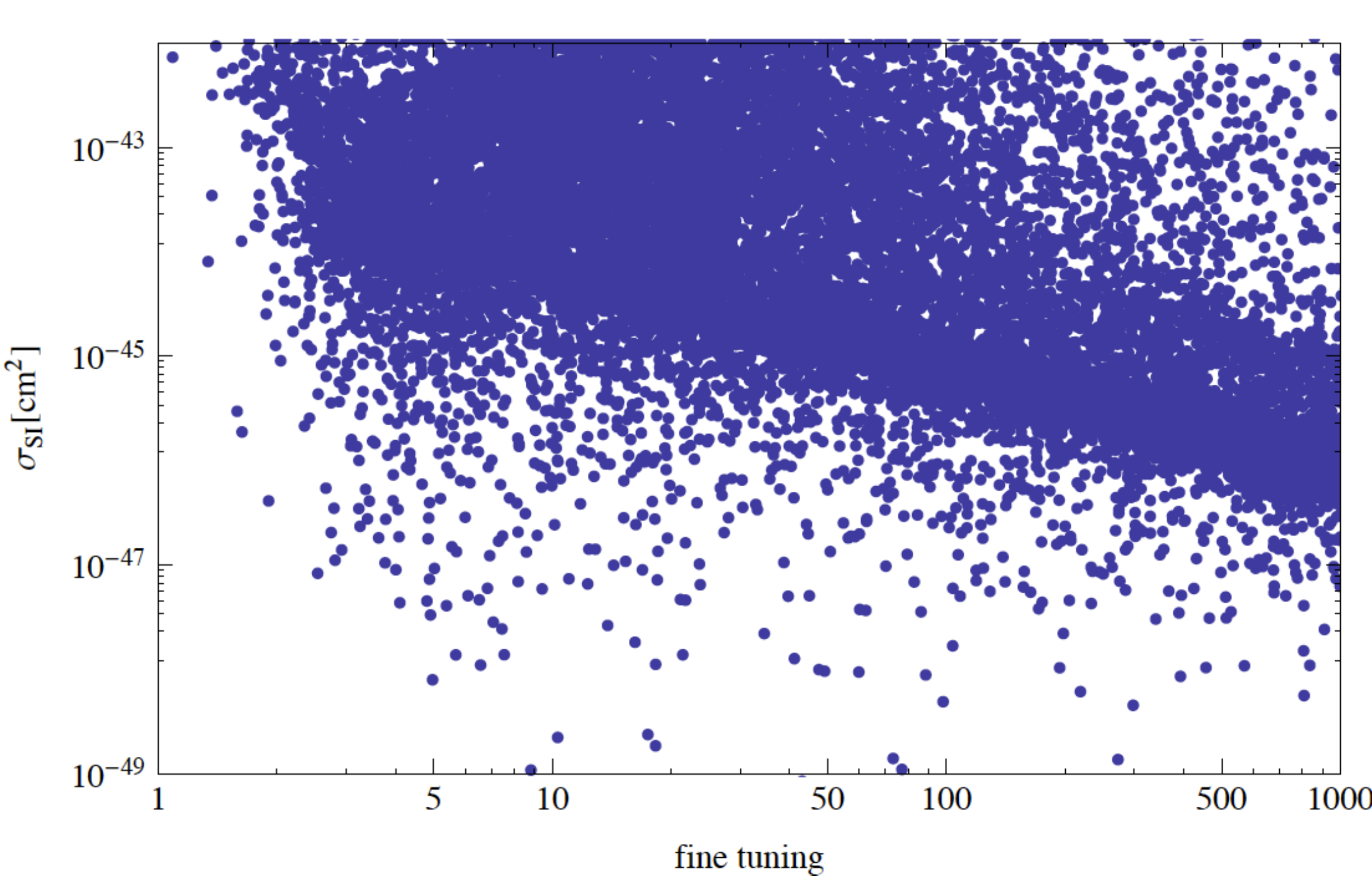}
\includegraphics[width=3.0in,height=2.1in]{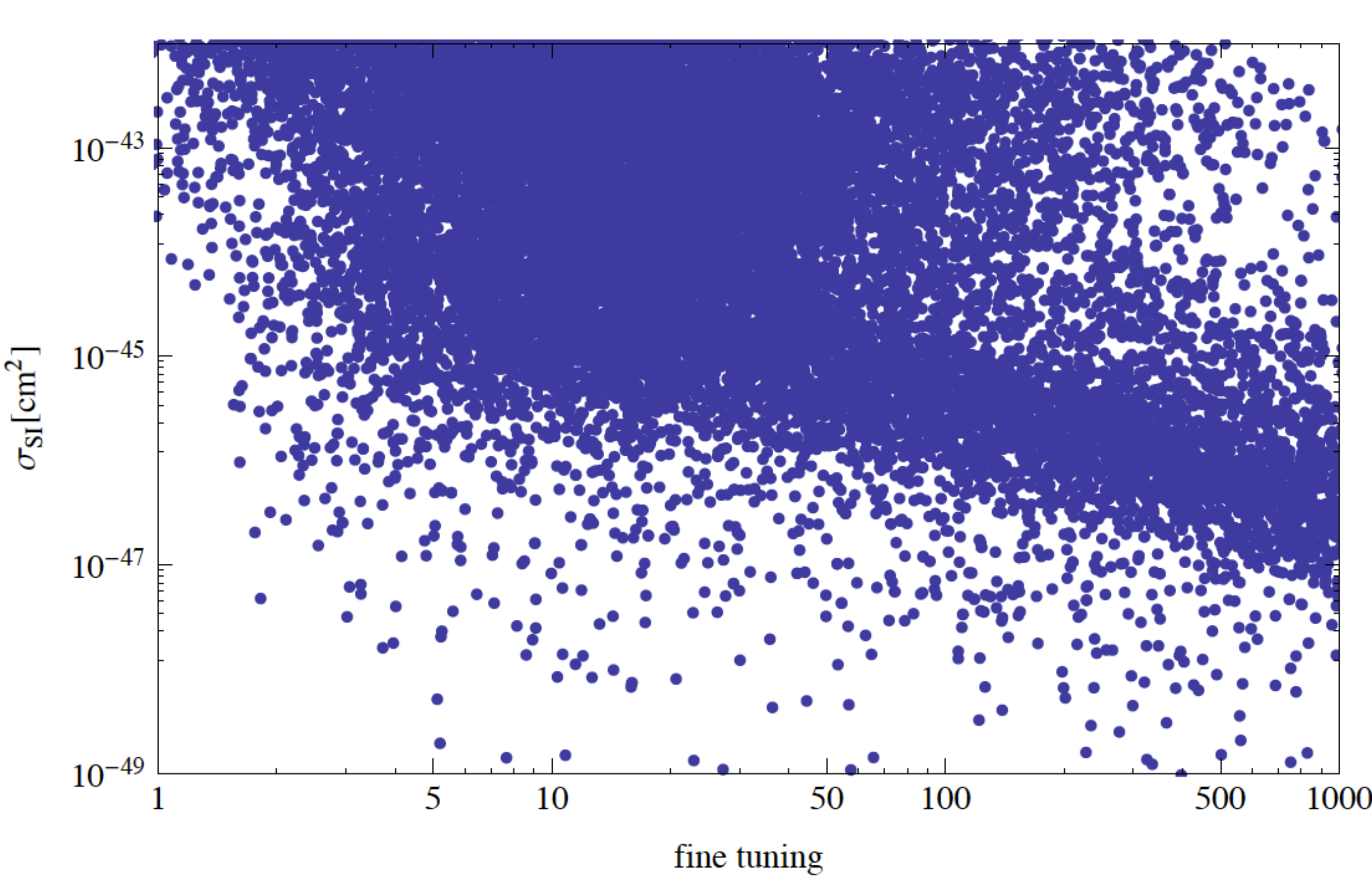}
\\
\includegraphics[width=3.0in,height=2.1in]{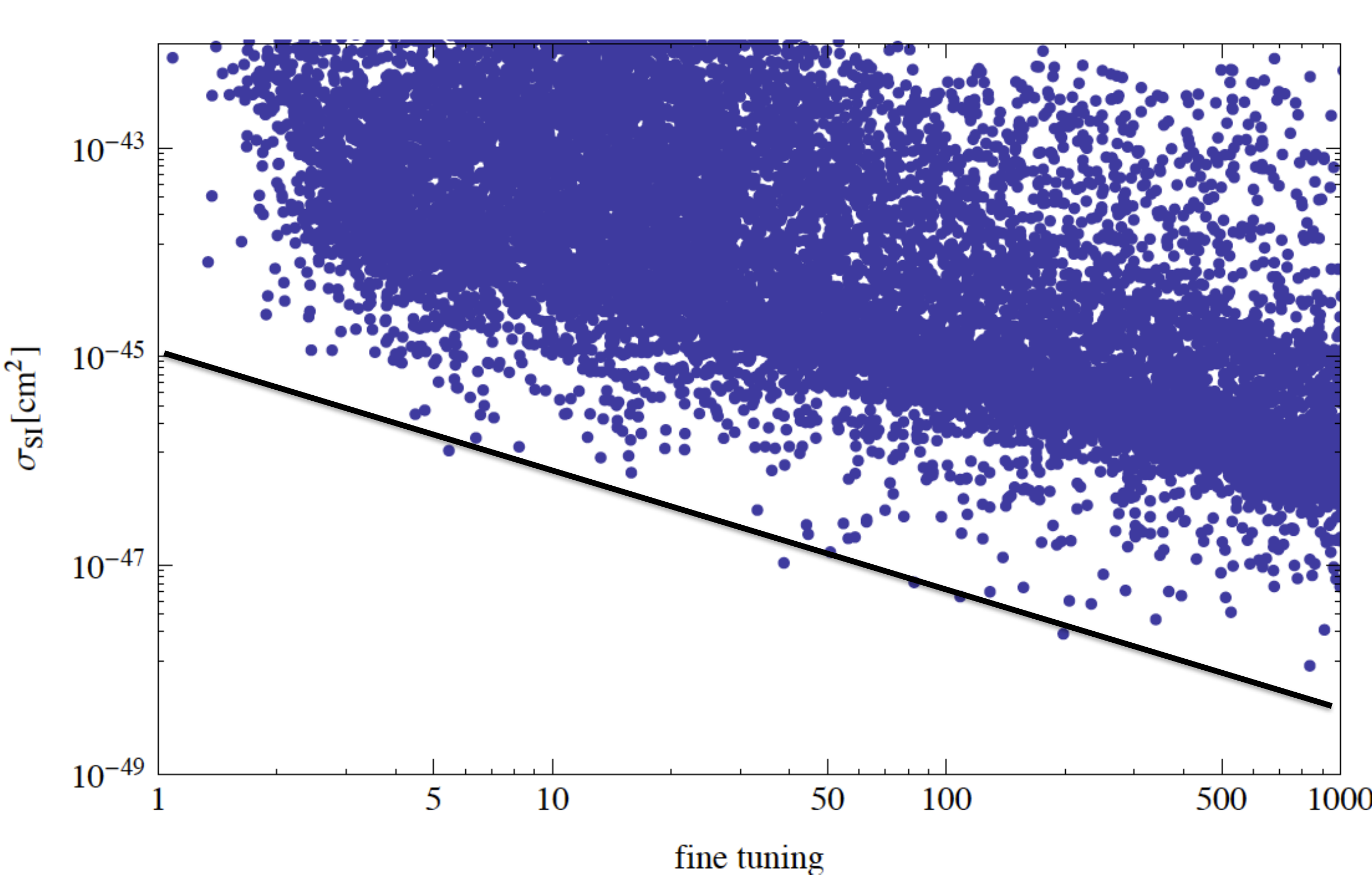}
\includegraphics[width=3.0in,height=2.1in]{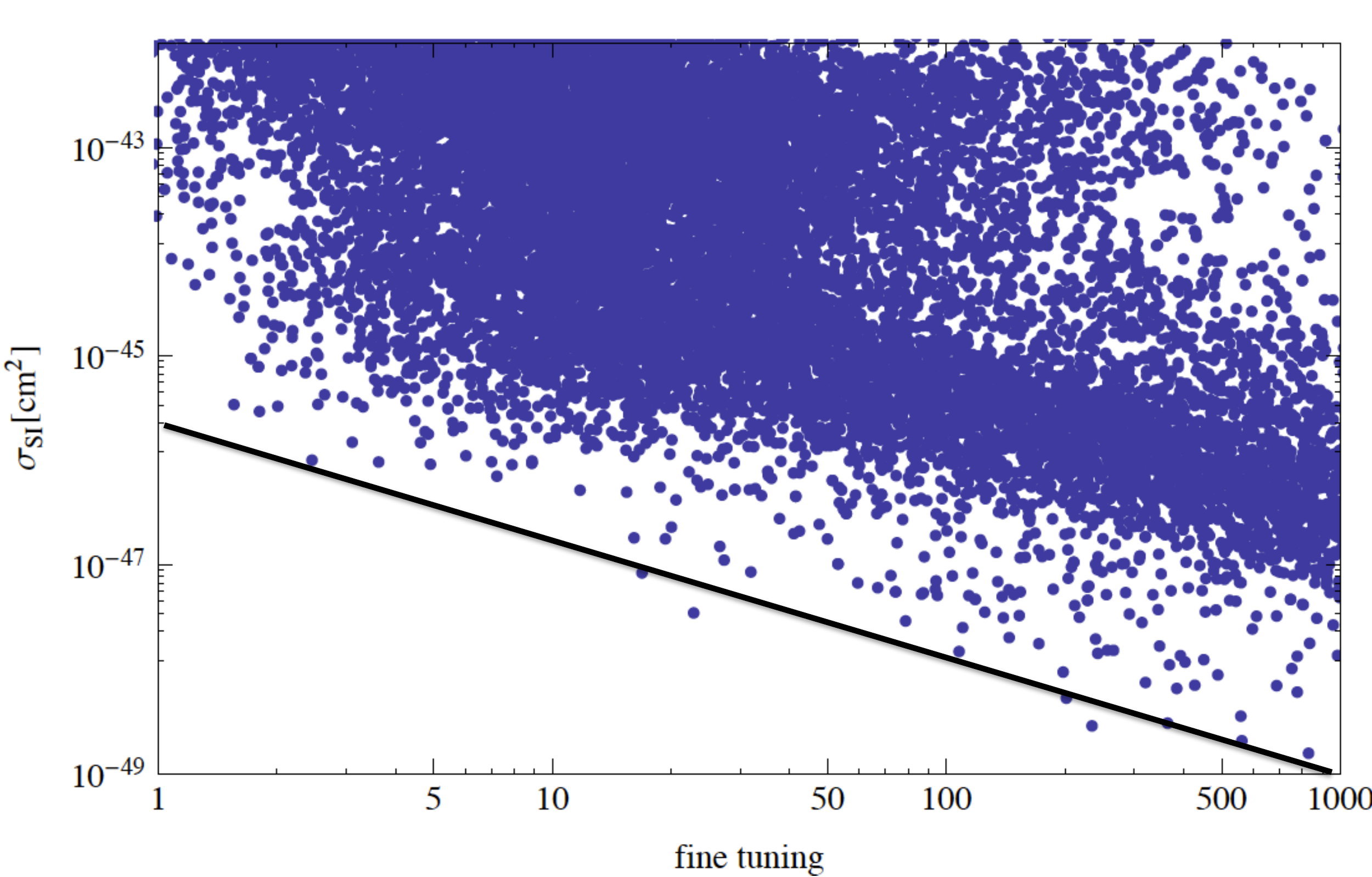}
\caption{Direct detection cross section vs. EWSB fine-tuning, in the NMSSM (left) and $\lambda$-SUSY (right). In the top row, all points are included; in the bottom row, points with accidental cancellations (Acc$>50$) are discarded. The lines represent approximate lower bounds from Eq.~\leqn{cs_bound}.}
\label{fig:csft}
\end{figure} 

\begin{figure}[t]
\centering
\includegraphics[width=3.0in,height=2.1in]{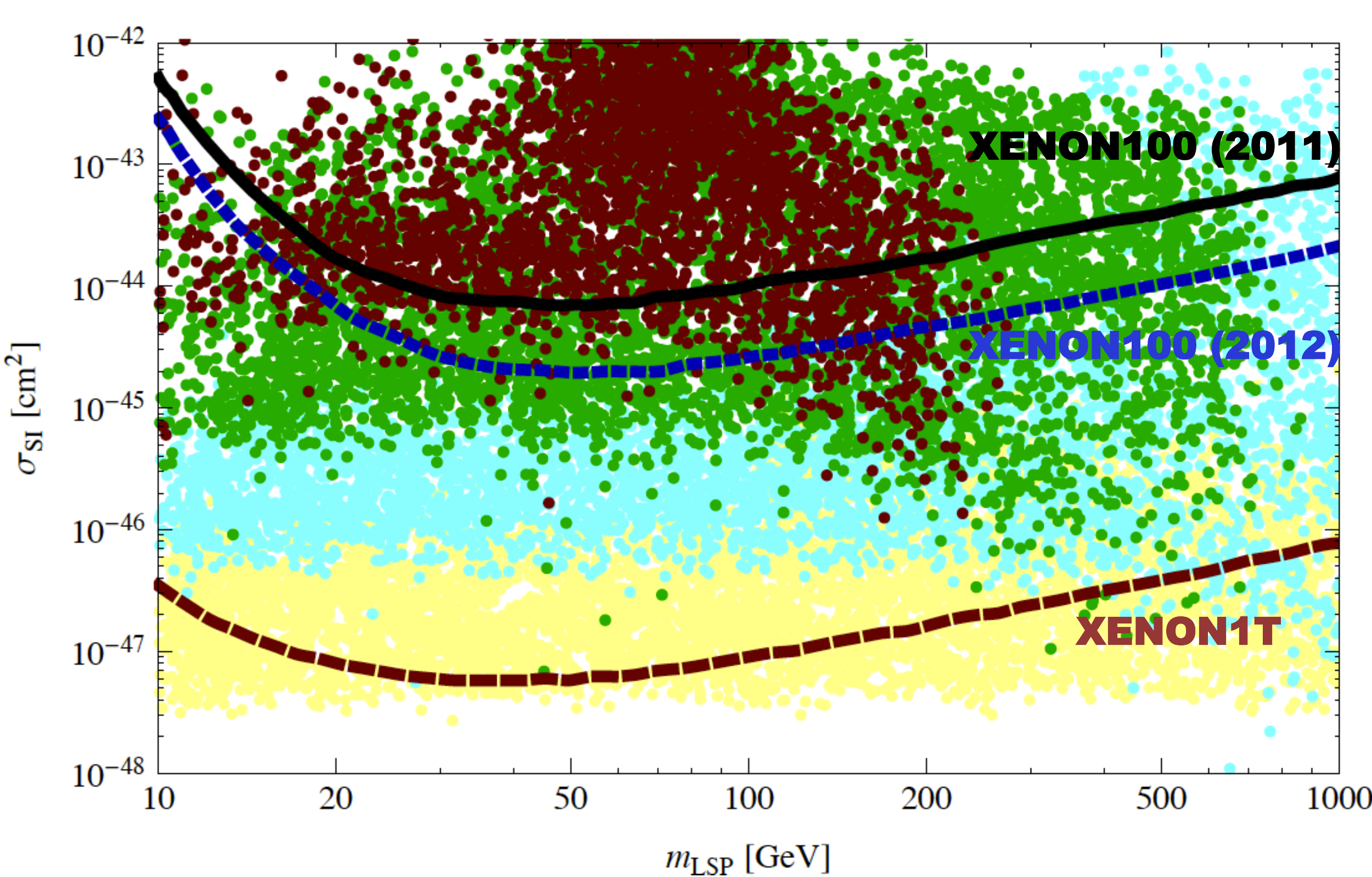}
\includegraphics[width=3.0in,height=2.1in]{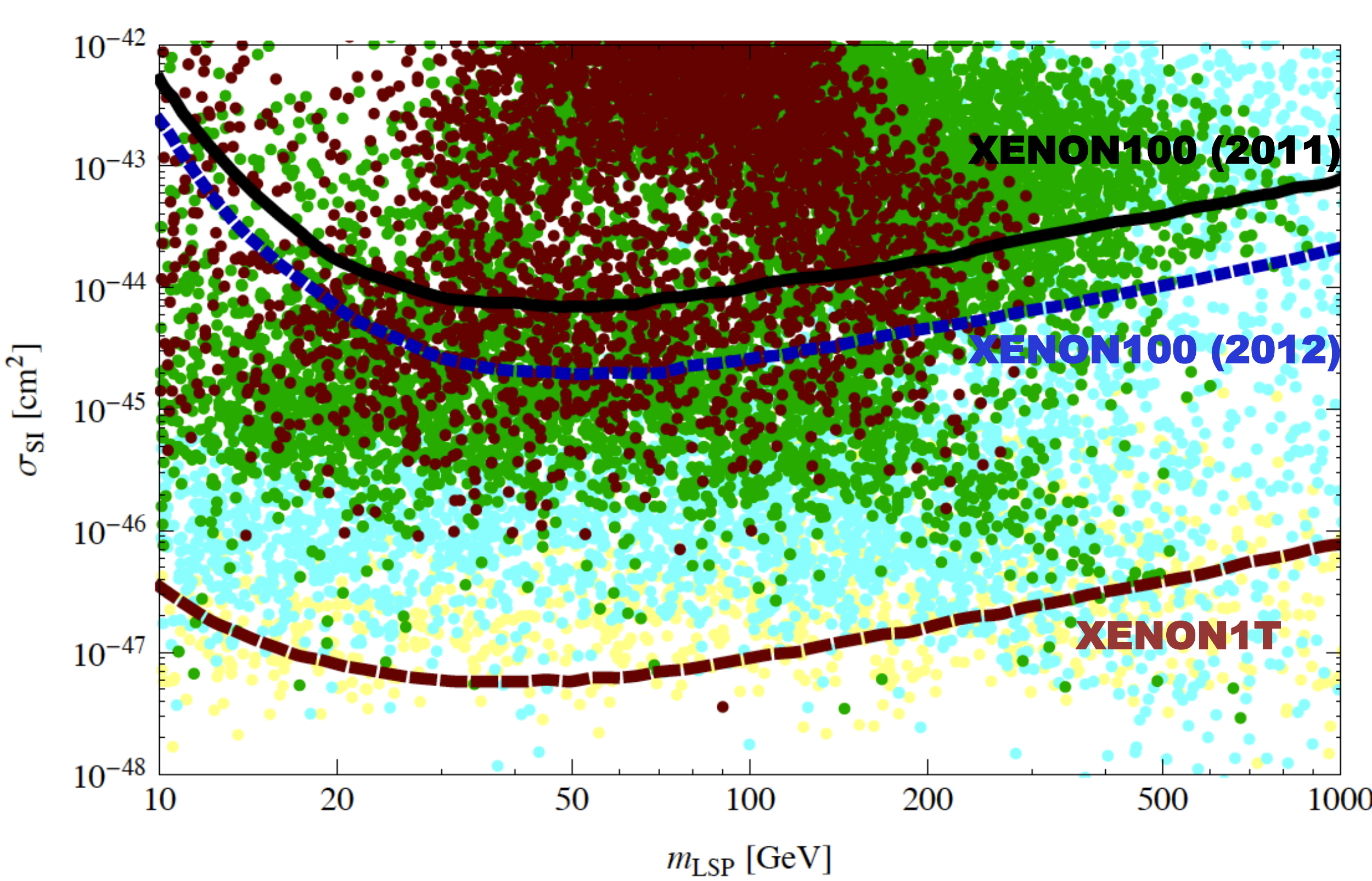}
\caption{Direct detection cross section vs. LSP mass, in the NMSSM (left) and $\lambda$-SUSY (right). Red, green, cyan and yellow points correspond to EWSB fine-tuning in the intervals $(0, 10)$; $[10, 100)$; $[100,1000)$, and $>1000$, respectively.  Points with accidental cancellations (Acc$>50$) are discarded. Lines denote XENON bounds and projections, as in Fig.~\ref{fig:MSSM}.}
\label{fig:csft_mass}
\end{figure} 

The most interesting correlation is between the direct detection cross section and EWSB fine-tuning. This is shown in Figs.~\ref{fig:csft} and~\ref{fig:csft_mass}. In these plots, we do not include points with Higgsino fraction in the LSP above $0.9$; the predominantly Higgsino case will be discussed separately below. The correlation found in the MSSM is preserved in the NMSSM and $\lambda$-SUSY: Points with lower direct detection cross sections have stronger fine-tuning, unless accidental cancellations occur. The physical origin of the correlation is the same as in the MSSM: suppressing direct detection cross section requires suppressing the Higgsino fraction, which can only happen (for a fixed LSP mass) by raising $\mu$, but this necessarily increases fine-tuning. It does not matter for this argument whether the LSP is gaugino, singlino or some combination of the two. The only caveat is the pure-singlino case with $\lambda\to 0$ and $\kappa\to 0$, where cross section can be suppressed with no fine-tuning price; however, as explained above, this region is not physically interesting. 

As in the MSSM, there is an approximate lower bound on direct detection cross section consistent with a given amount of fine-tuning. As a rough bound, we estimate from Figs.~\ref{fig:csft} that, in the absence of accidental cancellations,
\beq
\sigma_{\rm SI} \gsim \frac{10^{-45}~{\rm cm}^2}{\Delta}~({\rm NMSSM});~~~~\sigma_{\rm SI} \gsim \frac{10^{-46}~{\rm cm}^2}{\Delta}~(\lambda-{\rm SUSY}).
\eeq{cs_bound} 
The bound depends on the LSP mass, and the above estimates apply to the lightest LSP masses included in our scan, $m_{\rm LSP}=10$ GeV; the bound is higher for heavier LSPs. This is illustrated in Fig.~\ref{fig:csft_mass}, which also shows current and projected XENON bounds.  In the NMSSM, most points with fine-tuning better than 1/10 are already ruled out, although some remain.\footnote{Note that in a more constrained setup, such as in Ref.~\cite{Kang:2012sy} where unification and thermal relic density were assumed, stronger fine-tuning may be needed to satisfy the XENON100 bounds in the NMSSM.} Most of the remaining points lie either in the very low LSP mass region, where the XENON bound is weakened, or in the ``tail" of the low cross-section points in the $100-300$ GeV mass range. The ``tail" region, which has no counterpart in the MSSM ({\it cf.} Fig.~\ref{fig:MSSM}), arises from the region of parameter space with mostly-gaugino LSP and the Higgsino and singlino masses at roughly the same scale. The presence of the singlino lowers the Higgsino fraction in the LSP for the same value of $\mu$, resulting in lower cross section for the same level of fine-tuning. 

In contrast to the MSSM and the NMSSM, in $\lambda$-SUSY a large region of completely natural parameter space remains unconstrained by XENON100 (see right panel of Fig.~\ref{fig:csft_mass}). The reason for this is the parametric suppression of fine-tuning at large $\lambda$, discussed in Section~\ref{sec:FTsup}. This effect allows points with the same amount of fine-tuning to have a significantly higher $\mu$ than they would in the MSSM or the NMSSM. Higher $\mu$ suppresses the Higgsino fraction, which in turn lowers the cross section. Overall, we conclude that $\lambda$-SUSY allows significantly lower direct detection cross sections than either the MSSM or the NMSSM, without conflict with naturalness. 

When the LSP is predominantly Higgsino, the direct detection cross section in the MSSM can in principle be suppressed very significantly without paying a fine-tuning price~\cite{ftmssm}: Keeping $\mu\sim m_Z$ and raising $M_1\gg m_Z$, $M_2\gg m_Z$ does not introduce tree-level tuning, and the only limitation comes from the one-loop contributions of weak-inos to $\mhu^2$ and $\mhd^2$. This ``Higgsino loophole" can only be removed if thermal decoupling is assumed, in which case $\mu\gg M_Z$ is required to match the observed relic density. In the NMSSM, the region $\mu\sim m_Z$, $M_1\gg m_Z$, $M_2\gg m_Z$ does {\it not} generically produce a pure Higgsino LSP, because the singlino does not decouple: the singlino mass is of order $\mu$, and the singlino-Higgsino mixing terms are of order $v$, see Eq.~\leqn{neu_masses}. Since the mixed Higgsino-singlino LSP has a large direct detection cross section, one would naively expect that there is no ``Higgsino loophole" in the NMSSM. However, although the singlino is not parametrically decoupled, a modest hierarchy between $\lambda$ and $\kappa$ appears to be sufficient to suppress the singlino fraction in the LSP, and hence the cross section, while keeping $\mu\sim m_Z$. As in the MSSM, the neutralino annihilation cross section in this parameter region is too high to produce the observed dark matter abundance in the standard thermal decoupling scenario unless $\mu\sim$TeV, reintroducing fine-tuning. 
\begin{figure}[t]
\centering
\includegraphics[width=3.0in,height=2.1in]{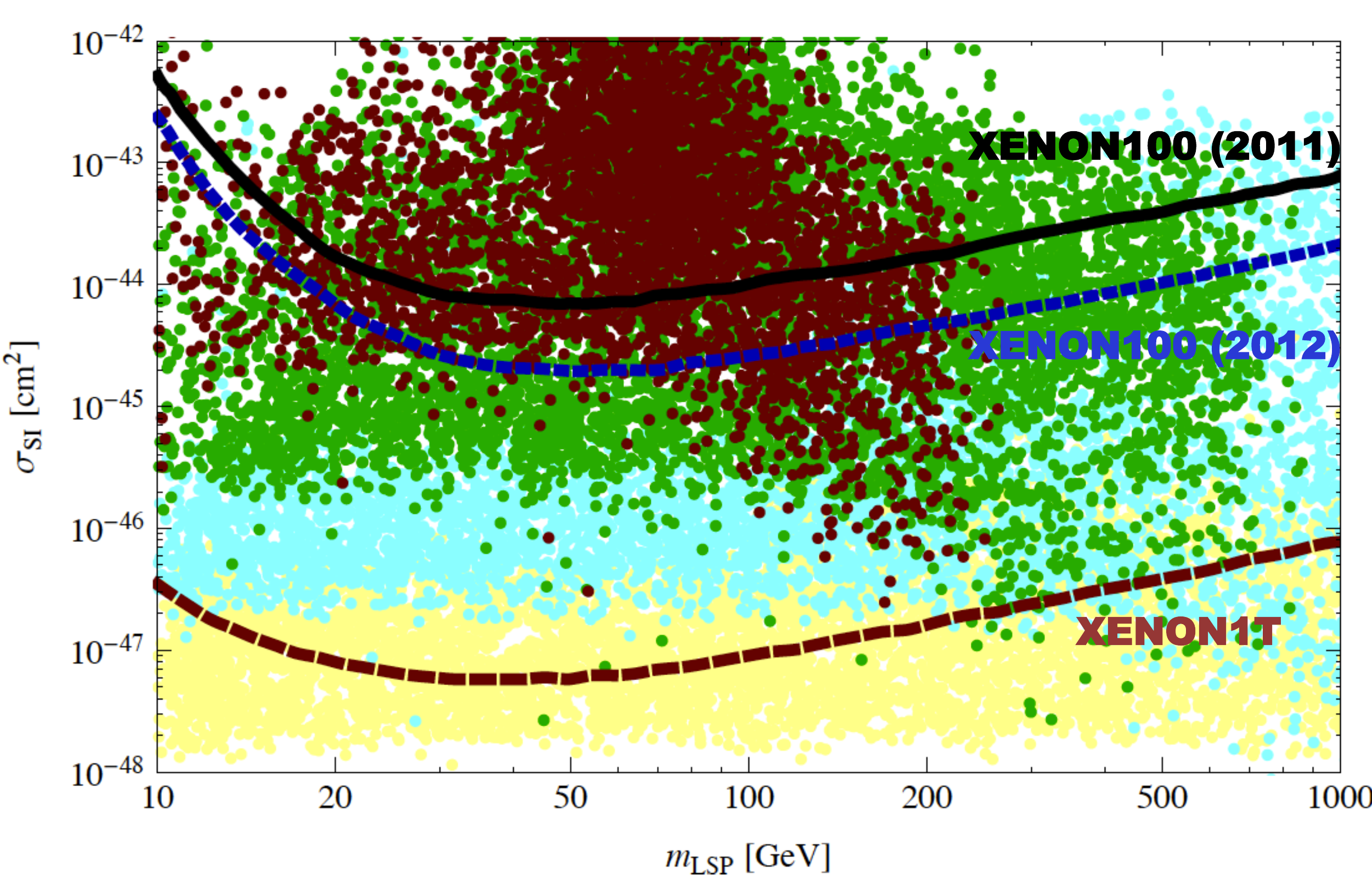}
\includegraphics[width=3.0in,height=2.1in]{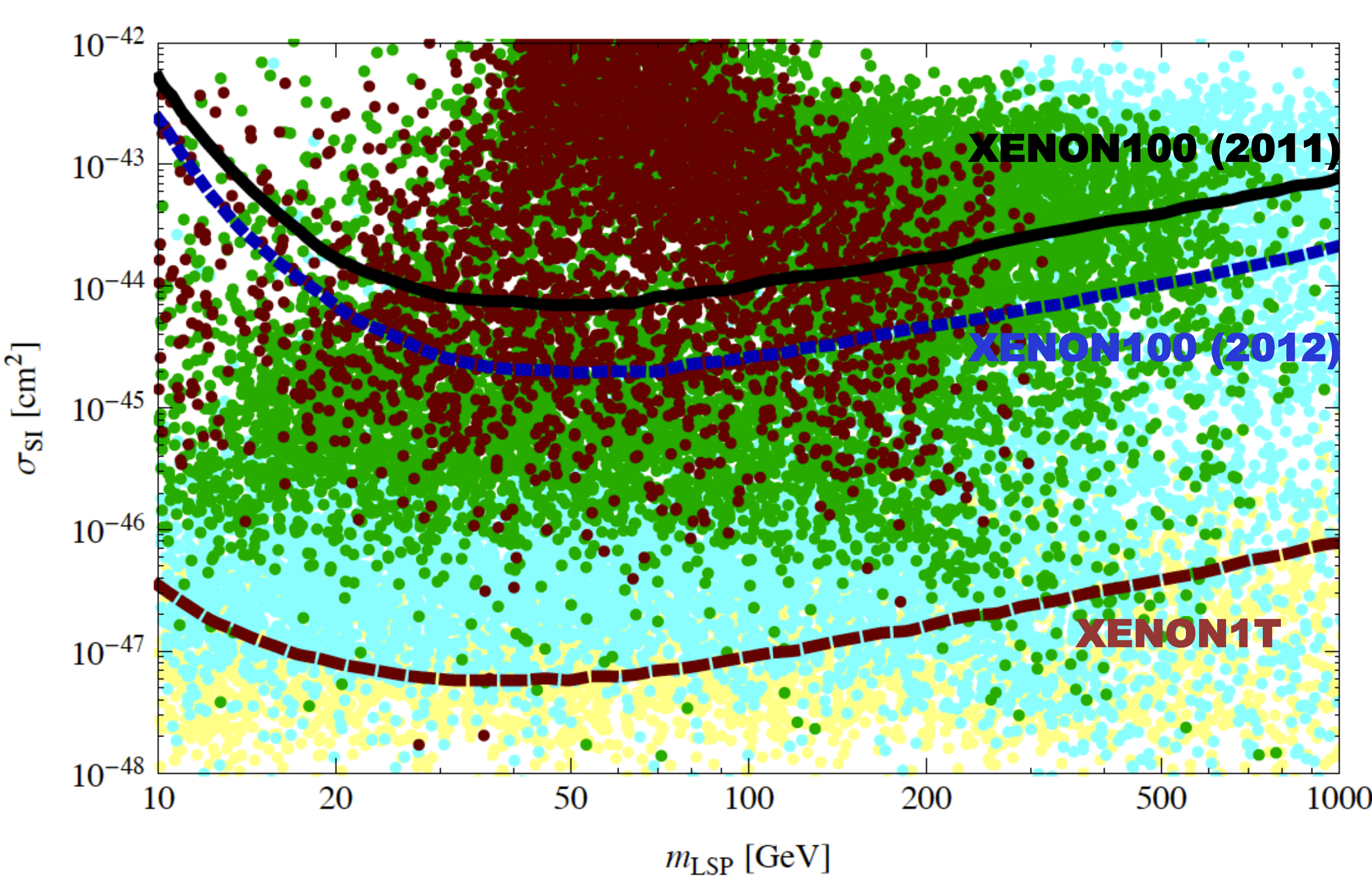}
\caption{Same as Fig.~\ref{fig:csft_mass}, but with $f_S=0.05$.}
\label{fig:csft_mass_lowfs}
\end{figure} 

It should be pointed out that, unlike in the MSSM, the CMS bounds on the CP-odd Higgs decays to $\tau$ pairs do not provide strong constraints in the NMSSM or $\lambda$-SUSY. This is because the CP-odd Higgs of the MSSM can now mix appreciably with the CP-odd singlet, suppressing the signal and thereby significantly weakening this bound\footnote{While recalculating the bounds from \cite{CMS:gya} to account for mixing with the CP-odd singlet, we have made the simplification of assuming that the signal comes entirely from gluon fusion production, and b-associated production is subdominant. This is accurate for low $\tan\beta$, which is our region of interest from naturalness considerations. Accounting for b-associated production would make the bound even less stringent.}.

As already mentioned in Section~\ref{sec:review}, there is significant theoretical uncertainty on nuclear form factors that enter the direct detection cross sections, with the strange quark form factor being especially important. Throughout the paper, we used $f_S=0.26$. An alternative lattice estimate is much lower, $f_S\sim 0.05$~\cite{lattice}. To illustrate the effect of this uncertainty on our conclusions, we repeated the scan with $f_S=0.05$. The results are shown in Fig.~\ref{fig:csft_mass_lowfs}. The correlation between the fine-tuning and direct detection cross section is unaffected. Numerical values of cross section are scaled down by an $\mathcal{O}$$(1)$ factor, so that more completely natural NMSSM points survive the XENON100 constraint. A better understanding of the nuclear physics is clearly very important for clean theoretical interpretation of the direct dark matter searches.    

\section{Conclusions}
\label{sec:conc}

In this paper, we continued our study, initiated in Ref.~\cite{ftmssm}, of the implications of dark matter direct detection experiments on supersymmetric models. In Ref.~\cite{ftmssm} we found that the MSSM, for ``generic" parameters, predicts a spin-independent elastic neutralino-nucleon scattering cross section of the order of $10^{-45}-10^{-44}$ cm$^2$ (depending somewhat on the poorly known strange quark form factor) or higher. Cross sections in this range are currently being probed by XENON100, which already places meaningful constraints. Suppressing the cross section below the ``generic" level in the MSSM requires (barring accidental cancellations) that the LSP be either a pure gaugino or a pure Higgsino. In the first case, lowering the direct detection cross section requires raising $\mu$, and therefore introducing fine-tuning in the EWSB. In the second case, requiring that the Higgsino be a thermal relic implies EWSB fine-tuning of about 1/500, independent of direct detection bounds.

The recent LHC discovery of a new particle with 125 GeV mass and properties consistent with the SM Higgs puts significant pressure on the MSSM, since fine-tuning of order $0.1$\% is required to accommodate it in this model. This motivates considering supersymmetric models with non-minimal Higgs sectors, where new contributions to the tree-level Higgs mass can easily arise. The simplest example is the NMSSM, where a single gauge-singlet superfield is added, and a 125 GeV Higgs can be incorporated with far less fine-tuning. The reduction of tuning is especially striking in the version of the NMSSM with strong doublet-singlet Higgs coupling, the so-called $\lambda$-SUSY. In this paper, we extended the analysis of Ref.~\cite{ftmssm} to the NMSSM and $\lambda$-SUSY. We found that the qualitative correlations between the dark matter direct detection cross section, the LSP composition, and the EWSB fine-tuning found in the MSSM essentially persist in these non-minimal models as well. Numerically, the minimal cross section allowed for the same level of EWSB fine-tuning is somewhat decreased in the NMSSM compared to the MSSM, and is further decreased, rather significantly, in $\lambda$-SUSY. We discussed the physical origin of these effects. We found that the current XENON100 cross section bounds are in mild tension with the MSSM and the NMSSM, excluding most points with fine-tuning of 1/10 or better, while large parts of completely natural parameter space are still allowed in $\lambda$-SUSY. 
 
As in Ref.~\cite{ftmssm}, we took a deliberately broad approach to the supersymmetric model parameter space. We do not assume any particular model of SUSY breaking; instead, we scan over unconstrained weak-scale parameters of each model. We do not impose the relic density constraints on the LSP.\footnote{Relic density constraints in the NMSSM are well known; see, for example, the review article~\cite{mainref}. Relic density constraints in $\lambda$-SUSY were considered in Ref.~\cite{Cao:2008un}.}  We also do not fully utilize the recent data concerning the 125 GeV Higgs candidate: We only require broad consistency of the Higgs spectrum with the data, demanding that a tree-level mass of at least one mostly-doublet CP-even Higgs be between 100 and 150 GeV. Imposing any combination of additional constraints would select a subspace of the broad parameter space studied here. Stronger conclusions can be obtained with such added constraints, but they would be less generally applicable. It would be interesting to perform such studies in the future.  

Looking ahead, we can anticipate further dramatic improvement in the experimental sensitivity of direct detection dark matter searches within a few years. Our results indicate that, if dark matter is supersymmetric, the searches will very likely be successful: Only supersymmetric models with sub-percent levels of EWSB fine-tuning, or accidental cancellations, will escape detection by experiments with sensitivity levels expected of, for example, XENON1T. In this paper, we showed that these statements apply not only in the MSSM but also in well-motivated non-minimal supersymmetric models. This underscores the importance of the continuing direct dark matter searches for fundamental physics. 

\vskip0.8cm
\noindent{\large \bf Acknowledgments} 
\vskip0.3cm

This research is supported by the U.S. National Science Foundation through grant PHY-0757868 and CAREER grant PHY-0844667. MP would like to acknowledge the hospitality of the Aspen Center for Physics, supported by the NSF Grant \#1066293. We would like to thank Roberto Franceschini, Aaron Pierce and David Sanford for useful discussions.

\begin{appendix}

\section{Fine -Tuning in $\lambda$-SUSY: an Analytic Example}

A simple analytic solution to the NMSSM scalar potential minimization conditions, Eqs.~\leqn{E1E3}, can be obtained in the absence of the two $A$ terms, $A_\lambda=A_\kappa=0$. Define $\tb\equiv \tan\beta=v_u/v_d$. The first two equations are a linear system with respect to $v^2$ and $\mu^2$. Solving this system provides expressions for these quantities in terms of $\tb$:
\beqa
\mu^2 &=& \mhu^2 \, \frac{1}{\tb^2-1} \left( -\tb^2 + \bar{m}_d^2  \right),\CR
v^2 &=& \frac{\mhu^2}{\lambda^2}\, \frac{\tb^2+1}{\tb^2-1}\left( 1-\bar{\kappa} \tb  - \bar{m}_d^2 (1-\bar{\kappa} \tb^{-1}) \right),
\eeqa{mu_v}
where $\bar{\kappa}\equiv \kappa/\lambda$, $\bar{m}_d^2 = \mhd^2/\mhu^2$, and terms of order $g^2/\lambda^2$ were neglected since we are interested in the large-$\lambda$ limit. The last of the equations~\leqn{E1E3} is then used to determine $\tb$:
\beq
\bar{m}_S^2 (\tb^2-1) \,+\, 2\bar{\kappa}^2  \left( -\tb^2 + \bar{m}_d^2 \right) \,+\,\left( 1-\bar{\kappa} \tb  - \bar{m}_d^2 (1-\bar{\kappa} \tb^{-1}) \right) (1+\tb^2-2\bar{\kappa} \tb) = 0\,,
\eeq{tb}
where $\bar{m}_S^2=m_S^2/\mhu^2$. This is a cubic algebraic equation on $\tb$. For Lagrangian parameters consistent with EWSB, it must have a real, positive solution, such that $\mu^2$ and $v^2$ are also real and positive, and the vacuum is stable ({\it i.e.}, the Higgs mass matrix does not have negative eigenvalues). Eqs.~\leqn{mu_v} and~\leqn{tb} together yield the equations~\leqn{A0mins} in the main text, and implicitly define the functions $f_1$ and $f_2$. (Explicit expressions can be obtained, but are not particularly illuminating.) For understanding fine-tuning at large $\lambda$, it suffices to observe that if all coefficients in Eq.~\leqn{tb} are order-one numbers, the resulting $\tb$ is also generically an order-one number. An example of Lagrangian parameters with consistent EWSB and the parameter hierarchy of interest to us is 
\beq
\lambda=2.0,~~\bar{\kappa}=0.5,~~\bar{m}_S^2=1.0,~~\bar{m}_d^2 = 0.5,~~\mhu^2<0\,.
\eeq{num_example}
At this point, $\tb\approx 2.12$, and we obtain $m_Z^2/\mhu^2\approx 0.05 $ and $m_Z^2/\mu^2\approx 0.04$, with no fine-tuning. (Note that this example is meant to be a simple illustration of the tuning suppression at large $\lambda$, and is not fully realistic: in particular, it does not contain a 125 GeV Higgs.)

\end{appendix}

\end{document}